# Distinct hydrologic response patterns and trends worldwide revealed by physics-embedded learning

Haoyu Ji[1], Yalan Song[1], Tadd Bindas[1], Chaopeng Shen[1,*], Yuan Yang[2], Ming Pan[2], Jiangtao Liu[1], Farshid Rahmani[1], Ather Abbas[3], Hylke Beck[3], Kathryn Lawson[1], Yoshihide Wada[3]

[1]Department of Civil and Environmental Engineering, The Pennsylvania State University, University Park, PA 16802
[2]Center for Western Weather and Water Extremes, Scripps Institution of Oceanography, University of California, San Diego, La Jolla, CA, USA
[3]King Abdullah University of Science and Technology (KAUST), Thuwal, Saudi Arabia
* Corresponding author: cshen@engr.psu.edu

## Abstract

To track rapid changes within our water sector, Global Water Models (GWMs) need to realistically represent hydrologic systems' response patterns — such as baseflow fraction — but are hindered by their limited ability to learn from data. Here we introduce a high-resolution physics-embedded big-data-trained model as a breakthrough in reliably capturing characteristic hydrologic response patterns ("*signatures*") and their shifts. By realistically representing the long-term water balance, the model revealed widespread shifts — up to ~20% over 20 years — in fundamental green-blue-water partitioning and baseflow ratios worldwide. Shifts in these response patterns, previously considered static, contributed to increasing flood risks in northern mid-latitudes, heightening water supply stresses in southern subtropical regions, and declining freshwater inputs to many European estuaries, all with ecological implications. With more accurate simulations at monthly and daily scales than current operational systems, this next-generation model resolves large, nonlinear seasonal runoff responses to rainfall ("*elasticity*") and streamflow flashiness in semi-arid and arid regions. These metrics highlight regions with management challenges due to large water supply variability and high climate sensitivity, but also provide tools to forecast seasonal water availability. This capability newly enables global-scale models to deliver reliable and locally relevant insights for water management.

## Introduction

While the global climate is driving more frequent extremes in precipitation and temperature[1], the terrestrial hydrologic system does not respond uniformly. Instead, the landscape modulates the impacts and feedback of these changes through complex and highly heterogeneous processes across space and time[2]. The landscape partitions precipitated water into evapotranspiration (ET) and runoff (composed of surface and groundwater contributions) and releases these fluxes at basin-specific rates. Because of strong storage threshold, memory and nonlinear effects[3–5], the landscape can translate the same amount of warming or changes in precipitation into either muted or disproportional changes in floods and droughts with high spatial heterogeneity[6,7]. To account for such nonlinear effects, Global Water Models (GWMs) such as WaterGAP[8], DBH[9], H08[10], LPJml[11], MATSIRO[12], and PCR-GLOBWB[13] were developed with integrated rainfall-runoff, river routing, and human water use processes to describe the terrestrial water cycle. GWMs are important tools referred to by organizations like the Intergovernmental Panel on Climate Change (IPCC) to project future changes in flood hazards, water availability, and societal resilience[14]. In a series of high-impact analysis work, GWMs have made great contributions to studying climate change effects[15] and water resource assessments[16].

The characteristic response patterns of a hydrologic system (often summarized as "hydrologic signatures") are interpretable, actionable yet challenging-to-model summaries that help stakeholders anticipate changes, identify challenges and manage risks[17,18]. Some examples of signatures include evaporation-precipitation ratio, baseflow-streamflow ratio, autocorrelation, runoff sensitivity to rainfall, and flashiness of the flow duration curve have strong implications for water management and aquatic ecosystem health[18]. These signatures control the quantity, timing, variability, temperature, and quality of freshwater exported downstream, which, in turn, exert first-order controls on the ecosystem composition of the subsequent water bodies. For example, drought-induced reductions in freshwater inputs to estuaries, controlled by precipitation-streamflow elasticity (discussed more below), can increase salinity for open estuaries and reduce salinity for intermittent ones[19]. The signatures can be highly valuable for decision makers. For example, understanding how the system partitions between ET and runoff can provide rapid quantification of available water resources; With accurate seasonal runoff sensitivity to rainfall, we can readily estimate the impact of forecasted multi-month droughts on summertime streamflows; Similarly, if we have a good grasp on the baseflow ratio, we can better estimate water quality and temperature[20]. However, a model that estimates these metrics needs to account for the hydrologic processes in all the (sometimes very large) upstream catchments, and must resolve systems' responses to inputs and recession behaviors at small timescales with high spatiotemporal precision. As we will show later, established GWMs can be challenged in these regards due to inadequate resolution and parameterization strategies.

Stakeholders worldwide have substantial urgent and unmet water prediction needs that current GWMs were not designed to address, due to the inherent tradeoffs between global coverage and high-quality local predictions (e.g., computational requirements). GWMs have so far mainly been tasked with describing large-scale, long-term change trends for entire climate zones, e.g., on long-term-average runoff on continental-scale large rivers[21,22]. It is not clear if these tasks have been fulfilled to the extent permitted by data, as GWMs have shown divergent behaviors in projecting future extremes[15] and precipitation-recharge responses[23]. Moreover, as a result of location-specific, scale-dependent hydrologic processes[24,25], coarse-resolution and inadequately-calibrated GWMs are not intended to be practical water management tools at local scales. For short-term tasks like flood and drought forecasts, GWMs' accuracy is often below operational requirements for daily or subseasonal forecasts, e.g., having daily Nash-Sutcliffe model efficiency coefficient (NSE) values lower than 0.5[26]. Until recently, stakeholders have either relied on detailed and costly area-specific model development and calibration for local tasks, or their needs were left unfulfilled, especially in developing nations. Thus, advancing global hydrological modeling to provide communities around the world with highly-accurate predictions should be a core concern of the scientific community.

With traditional modeling approaches, large observational datasets such as streamflow and soil moisture cannot readily benefit global predictions and communities. There are relatively few avenues for large observations to inform GWMs besides basic parameter calibration (e.g., with GloFAS[27]) and post-simulation bias correction (e.g., with WaterGAP[8] or GRFR[28]), which hinder the closure of the water balance. Even a rough calibration is hindered by the large computational demands of modeling at the global scale, and also suffers from the infamous issue of parameter nonuniqueness (or equifinality)[29]. Parameter regionalization, or generalizing parameters in space, remains a large and persistent challenge despite myriad proposed schemes[30]. Such methods cannot take advantage of the synergistic effects of large datasets, where observations from diverse sites jointly train one model to increase its robustness[31].

Recently, deep neural networks (NNs) have shown a formidable ability to learn from data and generate hydrological predictions, but have not yet benefited terrestrial water cycle assessment tasks. Purely data-driven networks like long short-term memory (LSTM)[32], transformers[33], and diffusion[34] have demonstrated success for simulating different hydrologic variables[35,36] and especially streamflow[37–40] at lumped small to meso basin scales. Nevertheless, their interpretability remains unsatisfactory: purely data-driven NNs do not have physical concepts like ET and baseflow when trained only on streamflow data for the calculation of the signatures. Furthermore, it is uncertain whether NN models optimized for daily NSE or Kling-Gupta Efficiency (KGE) can satisfactorily reproduce long-term trends in large rivers.

Leveraging the advantages of both process-based and deep learning models, a new opportunity has emerged to realistically describe the terrestrial water cycle. Physics-embedded machine learning (especially "differentiable") models[41] contain connected NNs and process-based equations which learn parameters or missing processes from data and are trained in a single step, enabling the complete tracing of inputs to outputs. Such models have shown comparable performance to LSTM while maintaining process interpretability and diagnosing untrained variables such as snowmelt, groundwater recharge, baseflow, and ET[42–46]. Due to the process-based components of the model, differentiable models generalize better than LSTM in data-scarce regions[47] and in representing extremes[48]. However, applying them for terrestrial water cycle assessment requires a performant and efficient river routing scheme to simulate major rivers, which necessitates improvements to our existing differentiable routing model[49]. Again, it remains uncertain if either the newer learnable models or established GWMs not trained on large data can accurately represent the spatial variability of hydrologic signatures and their temporal changes for continental-scale rivers and estuaries.

Here we demonstrate that by effectively learning from data at different scales, a physics-embedded differentiable hydrologic model can usher in transformational advances in the representation of global hydrologic response patterns, revealing previously-unrecognized hydrologic shifts occurring over the last twenty years. We answer these questions:

1. *How much has basic hydrologic partitioning, including the evapotranspiration-to-precipitation and baseflow-to-streamflow ratios, shifted in the past two decades worldwide according to data-trained models, and what are their implications for some important estuaries?*
2. *How do winter and summer streamflows worldwide respond to accumulated precipitation in the previous months, as characterized by seasonal "streamflow elasticity"?*
3. *Can big-data model training lead to the long-awaited step change in GWM performance — improving reliability for continental-scale impact assessment as well as increasing relevance to local stakeholders?*

Differentiable HBV+Muskingum Cunge routing (full version name δHBV2δMC2- Globe2-hydroDL, or δHBV2 for short in this paper) is a hybrid, multiscale model that can learn from thousands of sites and output hydrologic fluxes and states at high spatial resolution. Since its recent introduction in Song et al.[50], it has not been used to generate hydrologic insights or compared with GWMs yet. A neural network generates physical parameters for a differentiable implementation of the conceptual hydrological model Hydrologiska Byråns Vattenbalansavdelning (HBV)[51]. Rainfall-runoff processes are simulated with HBV's equations at small unit basins (MERIT network, median catchment size ~37 km$^2$) and then routed downstream by a differentiable Muskingum-Cunge (MC) model (Methods). 4,746 basins with catchment areas less than 50,000 km$^2$ are used for model training and testing. The evaluations are carried out on (i) 33 large global rivers with mixed anthropogenic influences (mixed-anthropogenic-impact rivers or "mixed rivers" for short) previously used in intercomparisons[52] of six models (GWM0-GWM5) in ISIMIP2a[53]; (ii) 28 other large rivers with less human impacts and a relatively minimal catchment area of 500,000 km$^2$ (natural rivers); and (iii) >4000 smaller basins with relatively continuous streamflow records.

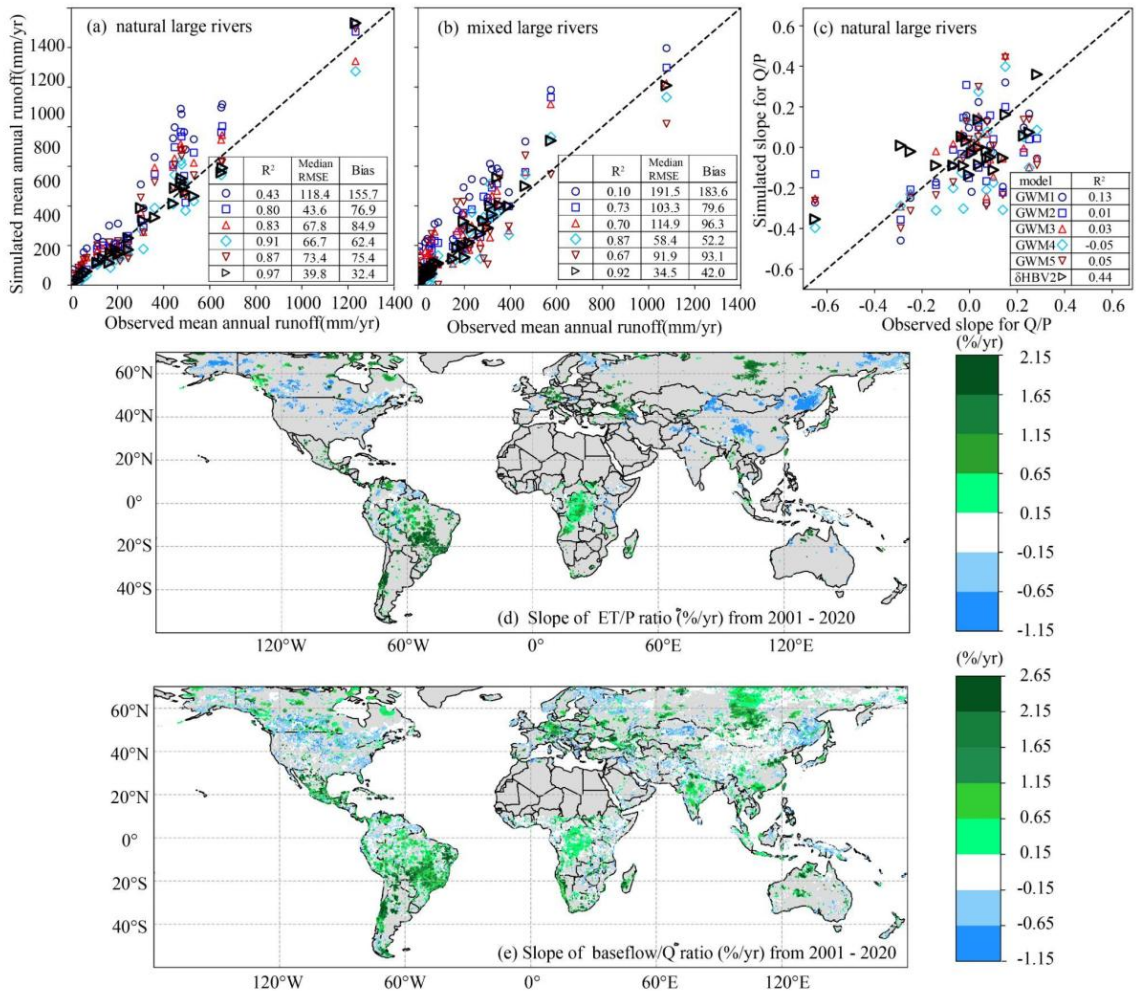

*Figure 1 | Global trends in water partitioning*
*(a–b) Simulated versus observed mean annual runoff (MAR) for large rivers with natural (a) and mixed (b) anthropogenic influences from 1981–2000. Insets report R², bias, and RMSE (calculate the interannual RMSE for each river and take the median among rivers) across models. All models conserve mass and apply no post-simulation bias correction. (c) Simulated vs. observed trends in runoff-to-precipitation ratio (Q/P) for natural rivers, with each symbol representing a model and river. (d–e) Spatial trends in ET/P and baseflow-to-streamflow ratio (baseflow/Q) from 2001–2020, shown as percent change per year in basins with statistically significant trends (Mann-Kendall, p < 0.05 colored).*

**Results and Discussion**

In the following, we first examine how basic hydrologic partitioning has changed between 2001 and 2020 and give practical examples of the implications for a number of US and European estuaries. To support the analysis in each case, we provide a benchmark against established GWMs in terms of matching various aspects of observations on large global rivers as well as smaller rivers.

Using simulation benchmarks from both large and small global rivers, we explain how the differentiable model can better capture such changes, which are challenging for established GWMs.

*Green-blue-water and baseflow-surface runoff partitioning*

Our assessment of terrestrial water partitioning relies on high-resolution simulations that can well capture water balances and their change trends. δHBV2 offers low bias for the long-term mean annual runoff (MAR) — it has a mean absolute bias of 32.4 mm/yr for natural rivers and 42.0 mm/yr for mixed rivers (Figure 1a-b, detailed value in Supplementary Table S1 & S2). These values are more than 20% lower than the biases of GWM4 (62.4 and 52.2 mm/yr) and 55% lower than those of GWM5 (75.4 and 93.1 mm/yr). δHBV2 had only 3 out of the 61 major-river basins (Conge, Xingu, Murtinho) with absolute biases over 100 mm/yr, fewer than GWM4 (13), GWM5 (17), and GWM1 (42). The high spatial $R^2$ values between MAR simulations and observations for δHBV2 (0.92 and 0.97 for mixed and natural basins. respectively) means most of the spatial variability in large-river MAR is explained by the model (Figure 1d-e). The interannual variability in streamflow was also well captured by δHBV2, which achieves the lowest average annual-scale root-mean-square error (RMSE) 34.5 mm/yr and 39.8 mm/yr for mixed and natural basins --- around 40% lower than GWM4 (Figure 1a-b). Importantly, δHBV2 is also the only model that achieved $R^2$>0.4 in describing the spatial variability of the temporal trends of the streamflow-to-precipitation ratio (Q/P, Figure 1c, an observable quantity related to ET/P via mass balance) for the natural rivers. Established GWMs tend to have $R^2$ values lower than 0.13 and large scattering in the estimated Q/P trends around the observed value, since such spatial heterogeneity in response patterns is challenging to grasp. Since these 61 large river stations were not used in training the model, δHBV2's high performance is due to the collective knowledge gained from the numerous small basins used for big-data training, which is crucial for capturing such shifts.

With refreshingly high accuracy and resolution, δHBV2 reveals significant trends in annual green-blue-water partitioning (quantified as evapotranspiration-to-precipitation ratio, ET/P) for many regions in the last 20 years (trends in Figure 1d; long-term average in Supplementary Figure S1). Blue water returns to the ocean (tied to flooding and irrigation), while green water (tied to plant water use and carbon/energy/nutrient cycles) exits as ET, and thus ET/P reflects the most fundamental partitioning of the terrestrial water cycle. In general, midlatitudes in North America and Asia and tropical areas like Central America and Papua New Guinea have seen increasing blue water while Central Europe and subtropical and midlatitude regions in South America have seen increasing green water. Some of these shifts are substantial — with a 1% change in this ratio per year, some regions have thus shifted 20% over the course of 20 years. These changes are correlated with trends in precipitation (Supplementary Figure S2) — where precipitation increases, blue water tends to increase, and vice versa, although the patterns do not fully match. This suggests that large-scale climate shifts affect water partitioning, and increasing rainfall can overflow storage thresholds to increase blue water[54].

The shifts in baseflow-to-streamflow ratio (baseflow/Q, trends in Figure 1e; long-term average in Supplementary Figure S3) have overlaps with those in blue-green-water, but are significantly more widespread. Due to the important role of groundwater discussed in Introduction, these shifts imply pervasive changes in stream temperature and water quality characteristics at decadal scale. Thus, baseflow ratio should not be treated static as done in most current studies[55]. It also exhibits regional clustering that were not noted before — a large region tends to have similar shifts, presumably reflecting decadal-scale trends in regional climate. The two ratios move correspondingly because both reflect increases in runoff and decreases in infiltrated or land-retained water. The processes of groundwater recharge and ET then compete for infiltrated water, as water in excess of the soil water-holding capacity moves to replenish deeper moisture and groundwater, which later becomes baseflow. ET/P changes are more muted than those of baseflow/Q, e.g., regions like India show noticeably rising baseflow/Q but little change in the ET/P. It can be explained that changes in precipitation and infiltration leave large imprints on recharge and thus baseflow, while the magnitude of ET responses may be limited by the ability of soil to hold water.

These shifts contribute to water excess and scarcity. Where blue-water fraction increases and baseflow ratio decreases in tandem (Figure 1 d-e), e.g., northeastern China, mid-latitude North America, and Papua New Guinea, there is a higher flooding potential, which has been documented in some studies[56,57]. Where ET/P increases substantially, it could be a sign of insufficient water for the ecosystem. The prominent shifts toward green water which occur in Germany, central Siberia, southern Brazil, central Chile, the Congo basin, and northern Australia are due to decline precipitation (Supplementary Figure S2) and suggest a disproportionate decrease in fast runoff in these regions. Because the baseflow process and hydrologic processes are scale-dependent, these fine-grained insights about baseflow and runoff need to be obtained from a data-trained high-resolution model.

As a direct consequence, we witness statistically significant trends in freshwater inputs to estuaries in the last two decades, with significant implications for ecosystems. The data-trained δHBV2 model identifies 10 estuaries (out of 55 analyzed) as having significant declining trends in mean annual freshwater inflows from 2001 to 2020, mostly along the North Sea coast of Germany and France (Figure 2b), which overlaps with the increasing ET/P and Baseflow/Q ratios. The declining trends are substantial — a few German sites decline more than 1.5% per year, amounting to a 30% decline in 20 years. The changes in ET/P ratio (Figure 1d) in Europe and US contributed to a larger than proportional impact. In contrast, US MidAtlantic estuaries have an increasing trend from 2001 to 2020. The trends for some of the mildly increasing/decreasing stations in the USA and France are not statistically significant (shown with thin marker borders in Figure 2b, but the regional clustering of such trends is clear. Freshwater declines can increase salinity and turbidity which can drastically alter macrofaunal communities including fish and invertebrates (including crustaceans, insects, and bivalves)[58,59]. Combined with global sea level rise[60], large changes in estuarine salinity and macrofauna were expected. Consistently, it was reported that macrofauna abundance in

German estuaries decreased by around 31% and the total biomass decreased by around 45% for approximately this period[61]. In that work, the changes were attributed entirely to sea level rise, but the decreasing freshwater inputs revealed here could have also played an important yet unacknowledged role.

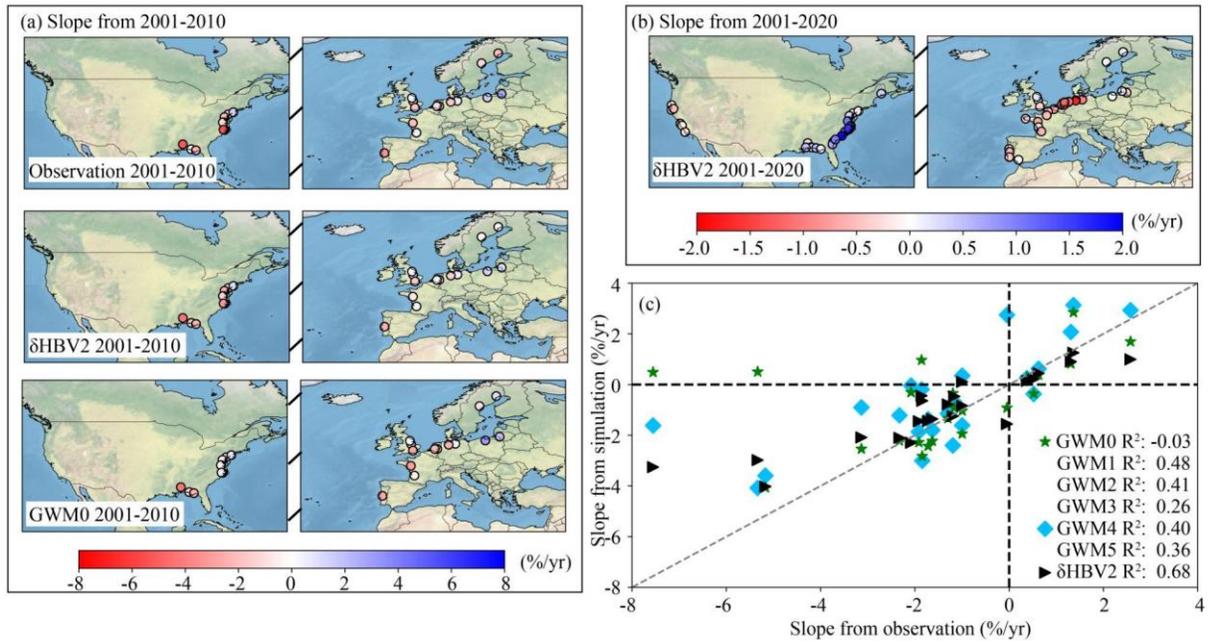

*Figure 2. | Estuary inflow trends and simulations by Global Water Models (GWMs).*
*(a) Simulated estuary inflow trends from δHBV2 and GWM0 compared to observed decadal-scale trends over the period of 2001 to 2010. (b) Simulated estuary inflow trends for δHBV2 from 2001 to 2020. (c) Scatter maps of observed estuary inflow trends and simulated trends from δHBV2 and GWMs from 2001 to 2010, where each point represents an estuary, with the symbol shape and color indicating the corresponding model (we only show GWM0 and GWM4 as examples). The $R^2$ value for each model is indicated in the legend.*

Perhaps partly attributable to model limitations, these above-mentioned declines in freshwater inputs to European estuaries (Figure 2c) have not previously been reported. δHBV2 describes noticeably different trends than established GWMs for these estuaries, as shown at sites where a comparison is possible (Figure 2a, 2001-2010, on sites where both observation data and GWM simulations are available). We found a stronger match of δHBV2 predictions with observed trends (reaching $R^2$ values around 0.68) compared to the GWMs (GWM0: -0.05, GWM4: 0.46, with other models ranging between them). GWM4, for example, overestimated the freshwater declines of European sites in 2001-2010 while not showing any trend for the US sites. Freshwater inputs show clear decadal-scale oscillations, as US Atlantic estuaries have notable declines between 2001-2010 and overall rising trends between 2001-2020 (Figure 2a), but GWM0 largely misses the downward swing (Figure 2a bottom row). Since δHBV2's conceptual model backbone is not fundamentally different from other models, the parameterization through big-data training appears to have greatly improved the sensitivity to decadal-scale changes, allowing us to identify or predict these trends.

*Seasonal streamflow response patterns*
To support assessing models' seasonal rainfall-runoff responses (called "elasticity"), we evaluate GWMs' abilities in capturing monthly flow fluctuations (Figure 3a), monthly runoff autocorrelation (Figure 3b) which relates to the recession behavior, and elasticity itself (Figure 3c). For the natural rivers, δHBV2 ranks 1st among all models with a median correlation of 0.89 at the monthly scale (Figure 3a), showing high success at capturing the seasonality of hydrology, although GWM0, which is enhanced by post-processing, achieves a slightly higher NSE. Unlike GMW0, δHBV2 does not impose a post-simulation bias correction and can ensure consistency for the internal hydrologic fluxes like ET, baseflow and recharge, meaning these variables are also indirectly informed by the learning process and consistent with alternative estimates. The strong representation for autocorrelation (Figure 3b for large natural rivers) suggests the model releases storage-dependent baseflow with the correct timing after conditioning by data, which is challenging for established GWMs. The established GWMs show substantial scattering in the arid regions for major rivers (high-range of the elasticity in Figure 3c), while the data-trained model stays closer to the 1-to-1 line. Overall, in all categories of the evaluated hydrologic signatures (one value per basin) related to rainfall responses, δHBV2 scores the highest except in ACF(1) (autocorrelation function with one month lag), where it places third with a margin of 0.01 (Supplementary Table S3). The difference in winter elasticity is smaller than in summer, presumably because the response in the wetter regime is faster and less nonlinear.

Examined regionally, δHBV2 tends to perform well in boreal or northern midlatitude rivers, but tends to be challenged in river basins where human water use is significant, e.g., Rio Grande de Santiago in the "northern dry" category, Cooper Creek in "southern dry" and the Columbia River, for which a large fraction of the catchment is arid (Figure 4). Other rivers with significant reservoirs, e.g., Danube and Missouri, can also be challenging. Among the 33 mixed rivers, 19 rivers have at least one GWM (excluding GWM0 — see discussion below) with a monthly Nash-Sutcliffe Efficiency (NSE) of 0.2 or above, and are regarded as meaningful for benchmarking (Figure 4a). Excluding GWM0, δHBV2 has the highest monthly NSE (median ~0.77) for 15 of these 19 rivers (Figure 4a).

The high-resolution δHBV2 enables the diagnosis of seasonal streamflow sensitivity to precipitation inputs (called "elasticity", $\varepsilon$), showing contrasting summer and winter $\varepsilon$ values that mostly complement each other (Figure 3d-e), with large summer $\varepsilon$ values in arid and semi-arid regions[62]. In searching for the aggregation lengths for the precipitation that generated the highest seasonal $\varepsilon$, we found it is 6 months for summer $\varepsilon$ and 3 months for winter $\varepsilon$. Except for central-western North America where there is some overlap, the high values for summer and winter $\varepsilon$ tend to be staggered (clustered in different regions), identifying these regions as "summer-responsive" or "winter-responsive". Summer $\varepsilon$ is the largest in arid and semi-arid regions, e.g., Sahel and Central Africa, central and southern South America, South Africa, central Asia, and northern China and Siberia (Figure 3d). In these regions, summer $\varepsilon$ is high because low flow is overwhelmingly driven by storage-dependent groundwater releases which are controlled by precipitation (rain or snow) accumulation in prior months. Winter $\varepsilon$ has a smaller range in values than summer $\varepsilon$, which is a possible outcome of more linear responses after thresholds are fulfilled. It is highest in central Asia and the northern Middle East, western Australia, the northern and eastern African coasts, eastern Brazil, and southern Patagonia (Figure 3e). In these regions, there is a more direct and immediate runoff response to winter rainfall. Such refined understanding is the first time it has been available by a high-resolution GWM.

For low-flow-dependent aquatic or riparian ecosystems, seasonal $\varepsilon$ is arguably more ecologically impactful than annual $\varepsilon$. High summer $\varepsilon$ regions are vulnerable to precipitation changes as reduction of seasonal precipitation there could have outsized impact on summer low flows. Hydrologically, this is because the reduced rainfall would not be sufficient to exceed some storage thresholds, resulting in disproportionate reductions in streamflows (blue water). Low-flow-dependent aquatic ecosystems in these regions could thus be sensitive to long-term changes in the precipitation regime or season-long droughts. However, high summer $\varepsilon$ also means stakeholders can use seasonal outlooks of precipitation to reliably predict streamflow (and inflow to the downstream water bodies) in the coming summer and prepare to intervene if possible. The $\varepsilon$ patterns appear noticeably different from some previous work that employed the Budyko curve for crude analysis[63], partly because here we analyze seasonal $\varepsilon$ rather than annual one, and partly a data-trained, high-resolution hydrologic model is available.

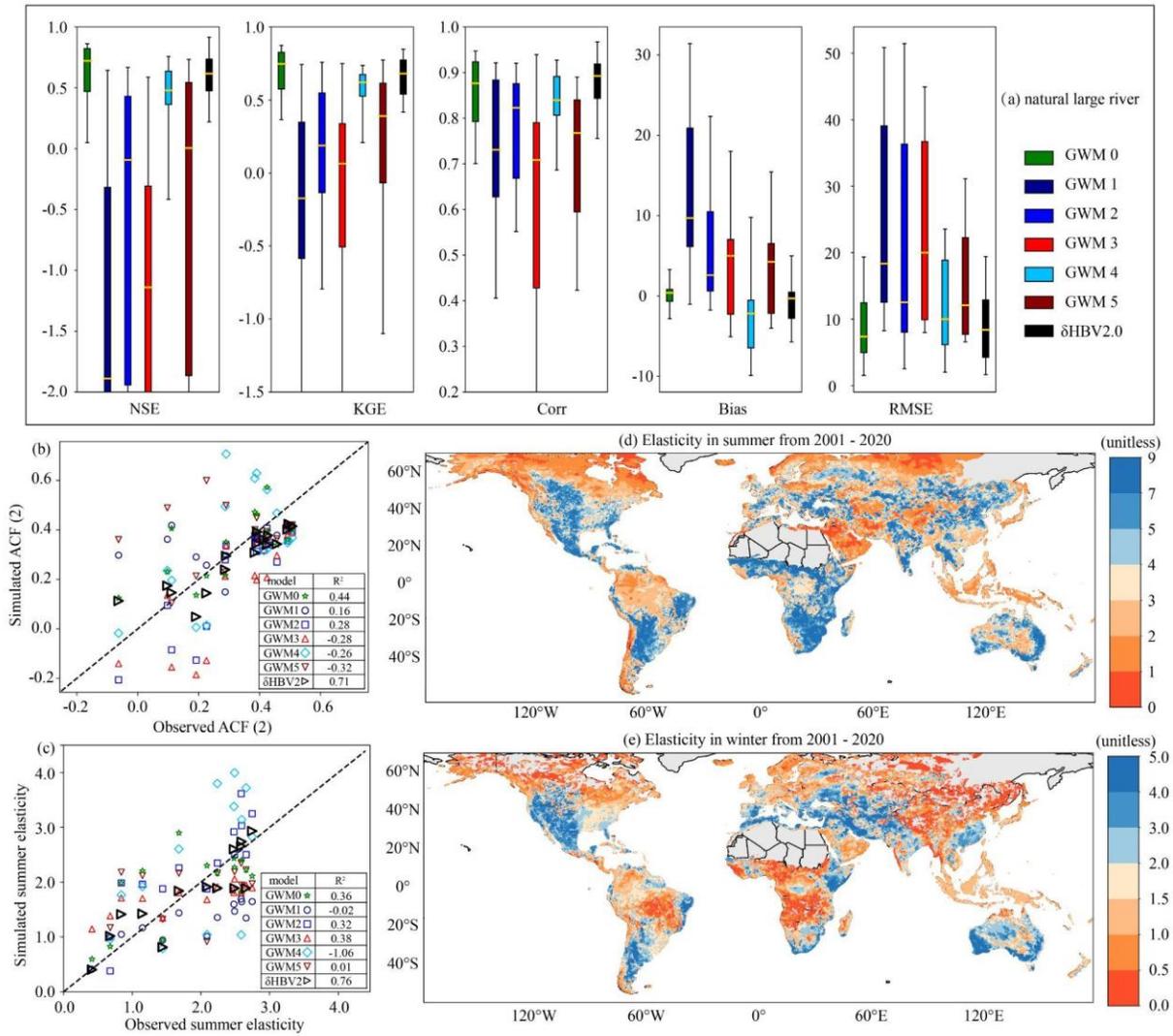

***Figure 3. | Monthly evaluations of GWMs and δHBV2-inferred seasonal elasticity.***
*(a) Monthly performance metrics—NSE, KGE, correlation, bias, and RMSE—for GWMs over large natural rivers. (b) Simulated vs. observed autocorrelation at a 2-month lag (ACF(2)) from 1981–2000. (c) Simulated vs. observed summer streamflow elasticity to 6-month precipitation (1981–2010). Each point in (b–c) represents a model–observation pair for one river. (d–e) Spatial patterns of streamflow elasticity to precipitation in summer and winter (2001–2020). Only regions with available data are shown.*

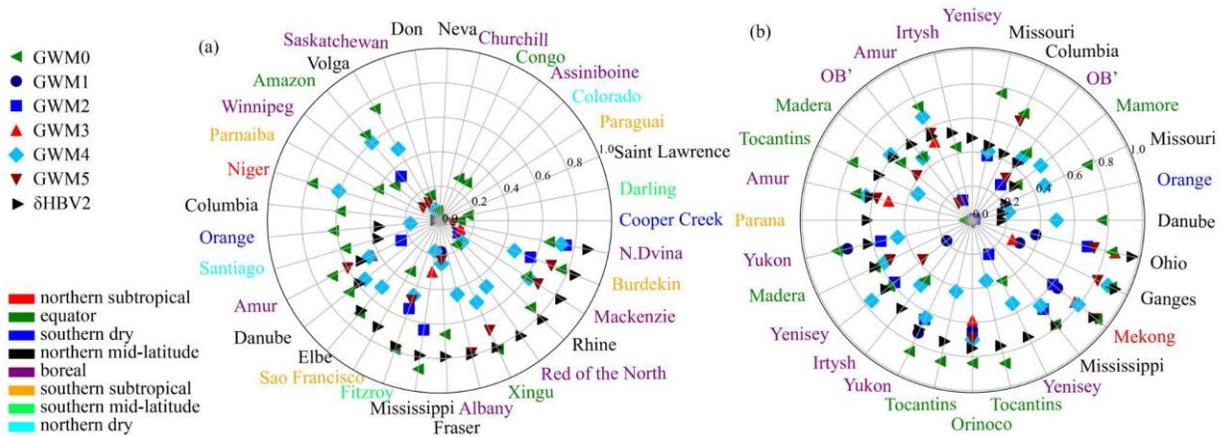

*Figure 4. | GWM performance across large global rivers. Monthly Nash-Sutcliffe Efficiency (NSE) scores (1981–2000) for global water models (GWMs) over (a) mixed-anthropogenic-impact and (b) natural basins. Rivers are arranged clockwise by δHBV2 performance (highest NSE on the outer ring, lowest at center). Marker shape and color represent different GWMs; river name color denotes biome type. The figure format and basin selection follow conventions from previous studies. GWM0 includes post-simulation bias correction; others do not. Amur, Columbia, Danube, Irtysh, Missouri, Madera, OB', Orange, Tocantins, Yenisey are repeated as the points in the right panel are taken from more upstream, natural gages of these rivers.*

## Daily streamflow variability and trends

To further validate the relevance of high-resolution δHBV2 for local water management, we evaluate its daily streamflow simulations in small-to-medium basins ($< 50,000$ km$^2$) against LSTM, lumped δHBV, and a widely-used operational global-scale product, GloFAS (Figure 5). We use several datasets (ds0-ds2, see Figure 5 caption) where comparisons are reasonable between different models. In a test spanning both training and testing basins, δHBV2 presents a high performance (median daily NSE of 0.63 for all basins and 0.53 for ungauged basins) significantly beyond that of GloFAS (median NSE of ~0.26) (Figure 5a). On another set of basins where comparison with GWMs is possible, δHBV2 noticeably leads GloFAS which, in turn, leads GWM0 (Supplementary Figure S4). The strong performance of δHBV2 stems from its big-data training for parameterization and generalization (i.e., leveraging data synergy[31]). δHBV2 shows a clear improvement compared to the original δHBV1.0 with its ability to resolve high-resolution heterogeneity, closely approaching LSTM in terms of KGE (median ~0.73) for the higher-performing basins (Figure 5b). This advantage against δHBV1.0 is more pronounced in data-rich regions, e.g., North America, where the subbasin-scale simulations can be better constrained by small training basins (Supplementary Table S4). There is still a slight gap between LSTM and δHBV2 in the lower half for daily KGEs (bottom left-hand side of Figure 5b), likely due to systematic forcing biases and unrepresented heavy water uses in some basins[48]. Both δHBV2 and LSTM thus represent state-of-the-art, allowing them to accurately capture the baseflow-surface runoff partitioning and flow variability. These results suggest that models with global coverage can finally be locally relevant for tasks such as flood forecasting and short-term water management.

The data-trained δHBV2 model showed large high-flow flashiness for arid and semi-arid regions and rather low flashiness for tropical rainforests with latitudes above 45 degrees (Figure 5c). High-flow flashiness is quantified by the slopes of the flow duration curve between the $1^{st}$ and $33^{rd}$ exceedance percentiles ($s_{FDC}$, with 1-33 percentile omitted for short) that are below -5 on the log scale, which corresponds to more than a 50 times difference between the $33^{rd}$ and $1^{st}$ percentile flows. Arid regions have large (the most negative) slopes due to very low baseflows and prevalent quick, heavy storms. The eastern US, Western and Central Europe, and southern China have moderate $s_{FDC}$ (around -3 ~ -1) as baseflow can be substantial. The tropical rainforests (Amazon, Congo, and throughout Pacific Asia) have the smallest $s_{FDC}$ as streamflow in these regions during even the dry season can be significant. While these results are not surprising, it is the first time they have been shown at the global scale using a high-resolution model.

Examining the changes in flashiness over time, we find widespread but spatially-mixed (statistically-significant) trends that do not easily match other spatial patterns studied here (Figure 5d). $s_{FDC}$ increases (becoming less negative) prominently in Mexico, western South America, and northern India, indicating a less variable distribution of streamflows in the last 20 years. However, central-western USA, southern Africa, the south fringe of the Sahara Desert, central Asia, and Ethiopia have seen large declining (more negative) trends, highlighting increasing streamflow variability. These changes are regional but can be substantial; we believe they may be caused by changes in precipitation intensities. Increasing flashiness poses the need for a greater storage capacity to ensure water supply resilience as well as control floods, although ecological consequences must also be considered. High resolution GWMs consistent with daily data are crucial for these applications.

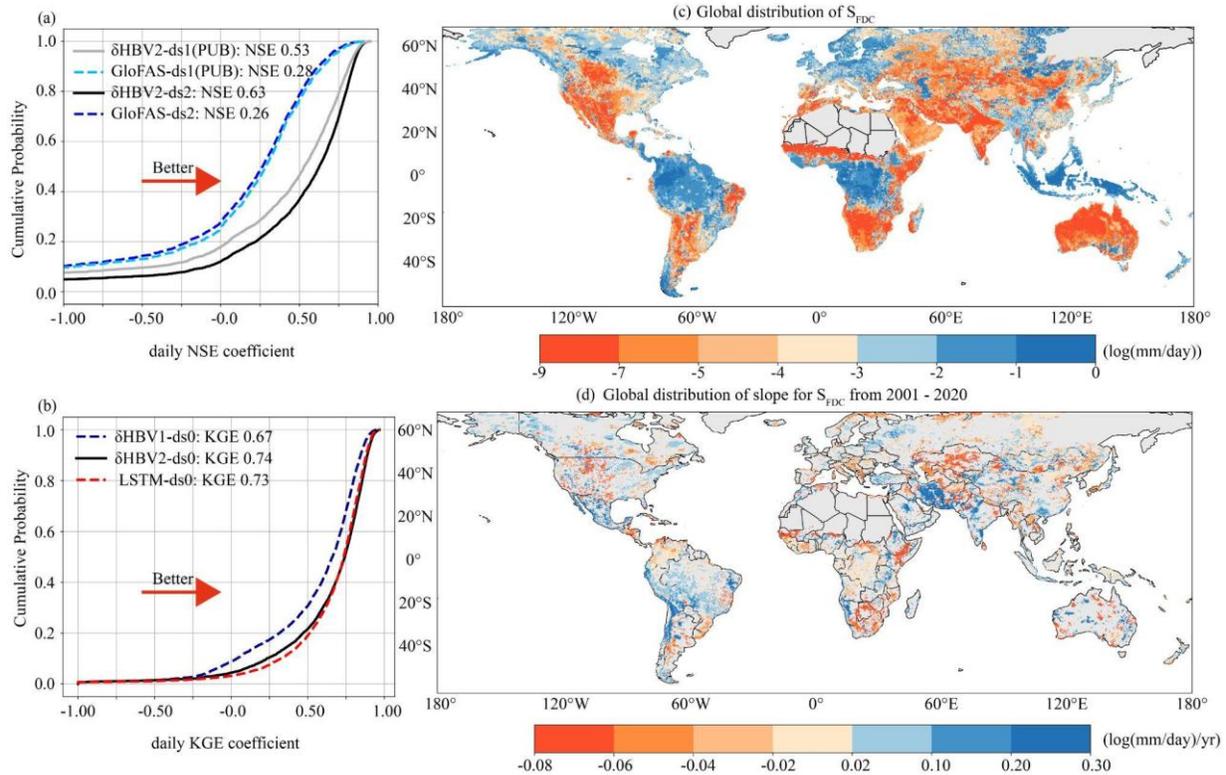

*Figure 5. | Flow Duration Curve (FDC) slopes and GWM performance comparisons.*
*(a) Cumulative distribution function (CDF) plots comparing δHBV2 and a widely-used operational global-scale product, GloFAS, on 5,558 basins (ds1) and prediction in ungauged basins (PUB, 2,509 basins not used in training, ds2). (b) CDF plots for temporal test results from δHBV1, δHBV2, and long short-term memory (LSTM, a purely data-driven model) on 4746 gages (ds0). ds0-ds2 are different sets of basins for comparing models as explained in Global Datasets section in Methods. (c) Global distribution map of Flow Duration Curve slope between 1% and 33% exceedance flow ($s_{FDC}$), used to indicate "flashiness" or the prevalence of sudden, heavy streamflows — a more negative slope indicates a high "flashiness". (d) Global distribution map of the temporal change rate (slope) of the slope $S_{FDC}$ (1% - 33%).*

### *Limitations*

Besides water withdrawals and very cold or very dry climates, certain downstream riverine and hydrological processes including large natural inland lakes, wetlands, and major reservoirs pose challenges for δHBV2. Large inland lakes and wetlands can store water, attenuate peak flows and sustain base flows, and induce floodplain water loss. They are not well simulated by the Muskingum-Cunge formulation. Examples include the Neva (Lake Ladoga), Paraguai (the Pantanal wetlands), Amazon (floodplains), Winnipeg (Lake Winnipeg), and Saint Lawrence (the Great Lakes) rivers. Due to the sizes of such storages, their impacts can even be noticeable at the monthly time scale. Additionally, flood-control dams, hydroelectric dams, and dams for other purposes such as irrigation and recreation each operate with distinct objectives, further contributing to the complexity of their impacts[64]. Combined with significant water withdrawals, it is challenging for the model to accurately capture such flow behaviors, which is only exacerbated in places with relatively few gages like Africa, e.g., the Niger and Congo rivers.

*Conclusions*

Driven by climatic shifts, the terrestrial water cycle is undergoing significant changes in quantity, timing, and hydrologic response patterns. Using our high-resolution, high-accuracy, physics-embedded model, we identified coherent and widespread shifts between 2001 and 2020 in fundamental water partitioning: both between evapotranspiration and runoff, and between surface and subsurface runoff. These changes are primarily nonlinear responses to precipitation variations. For example, North America and parts of Asia have seen increased blue water and surface runoff fractions—leading to higher flood potential, while the Southern Hemisphere, tropical rainforests, and parts of Europe have experienced increases in green water (i.e., evapotranspiration) and baseflow fraction—leading to reduced river flows. Consequently, freshwater inputs into some European estuaries have declined markedly, with associated ecological impacts. Arid and semi-arid regions already show high flow variability and runoff elasticity, making them especially vulnerable to future shifts in seasonal precipitation patterns. Some arid regions also see precipitation pattern changes leading to even higher flow variability. We provide a high-resolution map of the response patterns and changes to identify future challenges in water supplies and aquatic ecosystems.

While understanding hydrologic response patterns is helpful for water management, analyzing them and their changes requires a model that can accurately diagnose hydrologic fluxes with high spatiotemporal resolution. Traditional models often struggle to extract information from large datasets to characterize the landscape hydrologic responses, manifesting in the difficulty in describing the spatial distribution of the hydrologic signatures and their trends. Meanwhile, purely data-driven methods do not provide diagnostic variables or respect physical laws like mass conservation. Our approach—differentiable, physics-embedded learning—addresses both limitations. It offers, for the first time, a globally consistent, high-resolution, physically coherent picture of how hydrologic response patterns are shifting in response to climate, enabling local-scale fine-grained analysis and decision-making and revealing many previously unnoted changes described above. This modeling capability is essential for understanding and managing future water availability, aquatic ecosystem risks, and hydrologic resilience in a changing climate.

**Interest Statement**

CS has financial interests in HydroSapient, Inc., a company which could potentially benefit from the results of this research. This interest has been reviewed by the University in accordance with its Individual Conflict of Interest policy, for the purpose of maintaining the objectivity and the integrity of research at The Pennsylvania State University.

**Data and Methods**

*Global datasets*

δHBV2 was trained on a global dataset of daily observed streamflows, obtained from a recently-compiled global dataset by Abbas et al.[65], composed of various published resources as well as global and national databases. The published sources included the Global Runoff Data Centre (GRDC)[66], CAMELS series datasets for the United Kingdom[67], Brazil[68], Chile[69], Switzerland[70] and Germany[71], and LamaH-CE[72] for Austria, as well as the meteorological websites for the United States, Australia, France, Ireland, Poland, Finland, South Korea, and Japan. The initial dataset from Abbas et al.[65], comprising approximately 34,000 catchments, was narrowed down to 4,746 catchments with at least 95% data available from 1980 to 2020 (excluding data-sparse regions such as Africa) and with areas under 50,000 km². This choice was necessary in part due to the limited computational resources available for model training. The meteorological forcings dataset follows the strategy of Feng et al.[47] and includes the daily precipitation from Multi-Source Weighted-Ensemble Precipitation (MSWEP) V2.8[73], a product that includes global gage, satellite, and reanalysis data, as well as maximum and minimum daily temperatures from Multi-Source Weather (MSWX) V1[74], a product that bias-corrects and harmonizes meteorological data from atmospheric reanalyses and weather forecast models. The static attributes, including topography, climate patterns, land cover, and soil and geological characteristics, are derived from diverse sources and listed in Supplementary Table S5. To build the distributed model and discretize the global river flow, we used MERIT-Basins[64], a product that delineates global flowlines into discrete reaches and associates each reach with its predefined drainage area based on the 90-m MERIT-Hydro[76] digital elevation model (DEM) dataset. We used MERIT-Basins, with a median catchment size of 37 km², as our hydrological simulation unit.

δHBV2's MERIT-level resolution is much finer than GWMs' ~0.5 degree latitude-longitude grids. MERIT's flow network is delineated using high-resolution DEM, while GWMs are dependent on the 0.5-degree rectilinear boxes for 8-neighbor flow routing. As a result, there is discrepancy in the catchment area represented in the models, with MERIT having better topological correctness. Due to the low spatial resolution, GWMs have mostly been benchmarked only on large rivers at monthly or annual temporal scales. The first set of comparisons thus focus on long-term, annual and monthly evaluations on 33 large rivers benchmarked in the literature[52], which are under heavy human influences including reservoirs and water withdrawals. To evaluate the models under more natural conditions, we further added 28 large rivers (sometimes upstream gages of those among the first 33). The third set are the thousands of smaller gages where we compare δHBV2 to δHBV1, LSTM and GloFAS. GloFAS may still have catchment-area discrepancy with our model, and we apply a filter to ensure this discrepancy among all models compared must be within 20% of that the gage-registered value. Therefore, the gages on which we can compare δHBV2, δHBV1 and LSTM are different from those on which we can compare δHBV2 and GloFAS. We used several different datasets (*ds0-ds2*) for comparing models due to this filter. *ds0* contains 4746 basins for temporal test, with 1980-2000 as the training period and 2001 - 2015 as the testing period. *ds1* contains 2,509 basins for spatial test (or prediction in ungauged basins, PUB): they are not used in training by either δHBV2 or GloFAS, with 1980-2010 as the testing period. *ds2* (all basins) contains 5,558 basins, which is the union of *ds0* and *ds1*, while dropping some basins that failed the catchment area filter, to compare δHBV2 against GloFAS.

*Models*
As an overview, we used two differentiable models for global streamflow prediction in this work. Both models used connected neural networks (NNs) to generate physical parameters for process-based HBV models. The first differentiable model, as presented in Feng et al.[42] and Feng et al.[47] is a basin-lumped model (δHBV1.0) with inputs and parameters defined for the entire catchment upstream of each prediction gage. The second model is a high-resolution model defined on the MERIT unit basins, with a differentiable implementation of Muskingum-Cunge (δMC) as the routing method and flow aggregated to the downstream gage for comparison with the discharge observations. This model is called δHBV2.0δMC-Globe-hydroDL, or δHBV2 for short. Both models support saving their internal states (e.g., physical states and LSTM hidden states) to files for future restart or continued simulations. The lumped version of δHBV can only provide a point prediction at a basin's outlet, while the high-resolution model can provide hydrologic predictions at all locations along the seamless MERIT river network, with considering the spatial heterogeneity of inputs.

*Basin-lumped differentiable model (δHBV1.0)*
The original basin-lumped differentiable model, δHBV1.0, connects an NN, which supports regionalized parameterization, to a conceptual model that uses basin-averaged inputs to predict streamflow at the catchment outlet. Here, we connect a Long Short-Term Memory (LSTM) network to the Hydrologiska Byråns Vattenbalansavdelning (HBV) conceptual rainfall-runoff model[51,77]. δHBV can be described succinctly as:

$$\theta = LSTM(x_b^{l:t}, A_b) \tag{1}$$

$$q_b^{l:t} = HBV(\theta, x_b^{l:t}) \tag{2}$$

where $\theta$ represents the physical parameters of HBV. $A_b$ represents the static attributes for each basin, while $x_b$ represents meteorological forcings used to drive ("force") the HBV model, including precipitation, mean temperature, and potential evaporation. Components of $A_b$ and $x_b$ are listed in Supplementary Table S5. $q_b$ and $q_b^*$ are the simulated and observed streamflows, respectively, in units of mm/day. $q_b$ can be converted to $Q_b$ in units of m³/s using the basin area. Subscript $b$ denotes the particular basin for the basin-averaged inputs and outputs, which are "lumped" to contain the entire upstream catchment.

The structure and hyperparameters of LSTM can be found in Supplementary Text S1. It is a sequence-to-sequence neural network capable of processing input time series and capturing both long-term and short-term tendencies through its memory cell states[32], then producing target time series (e.g., physical parameters of hydrological variables). While many other architectures have been attempted, so far it is still challenging to surpass LSTM for deterministic prediction tasks in hydrology[78]. It can generate either static or dynamic parameters in HBV, i.e. the whole time series can be used as daily-varying parameters. For this work, we evaluated δHBV1.0 with only static parameters. The parameters obtained from the last day of LSTM outputs are used as static parameters.

The structure and physical parameters of the HBV model are provided in Supplementary Table S6. HBV can simulate snow accumulation and melt, soil moisture, evapotranspiration, and runoff generation. HBV as employed in this work assumes 16 parallel components, meant to mimic the hydrologic response units that capture subbasin-scale heterogeneity. δHBV1.0 and the subsequent differentiable models are implemented on PyTorch[79], a Python library that supports automatic differentiation (AD) and highly parallel execution on Graphical Processing Units (GPUs). The NN for parameterization and the re-implemented HBV equations are trained in a single step to allow gradient tracking and the use of gradient descent (Eq. 3). The overall framework can be trained simultaneously over many basins to obtain a regionalized and generalized mapping from environmental variables (including the meteorological forcings and static attributes) to HBV parameters. Once trained, it can easily generate parameter maps over untrained basins. The goal of the training process is to minimize the loss function, which is based on the root-mean-square error calculated between each simulation and its corresponding actual observation:

$$Loss_{RMSE} = (1.0 - \alpha)\sqrt{\frac{\sum_{b=1}^{B}\sum_{t=1}^{T}(q_b^t - q_b^{*t})^2}{B*T}} + \alpha_l\sqrt{\frac{\sum_{b=1}^{B}\sum_{t=1}^{T}(log(q_b) - log(q_b^*))^2}{B*T}} \quad (3)$$

where $B$ is the total number of gage basins in a training batch, and $T$ is the number of the simulation days for the batch (time length). The mean square error of the log-transformed streamflow is added in the loss function to better consider the low flows in the training. $\alpha$ is a weight used to balance the performance trade-off between the high flow and low flow, where a large $\alpha$ will emphasize low flows in the training (Eq. 3). We adopted the same weight as used in Feng et al.[42], which was 0.25.

*High-resolution, multiscale differentiable model (δHBV2.0)*
δHBV2 is a high-resolution, multiscale differentiable model which has been adjusted to better capture the heterogeneity of large basins and address hydrologic scale discrepancies. This model takes the inputs and makes predictions at the resolution of small MERIT unit basins (median basin area is 37 km²). The connected neural network (NN) estimates HBV parameters at the MERIT unit basin level, where the HBV rainfall-runoff model is also run. The unit-basin-level runoff is aggregated to the outlet of the gage basin to be compared to the gage data. The NN provides time-variant parameters to compensate for missing processes, e.g., vegetation dynamics and deep water storage and return flow.

To summarize, an LSTM network and a Multilayer perceptron (MLP) are used to generate dynamic (daily time-variant) and static HBV parameters, respectively:

$$\theta_{d,m}^{1:t} = LSTM(x_m^{1:t}, A_m) \quad (4)$$

$$\theta_{s,m} = MLP(A_m) \quad (5)$$

where $x_m$ and $A_m$ are the same meteorological forcings and geographic attributes used as inputs in δHBV1.0, where subscript $m$ denotes the based-averaged variable for MERIT unit basins, and subscripts $s$ or $d$ respectively denote static or dynamic parameters. To suppress overfitting, we only chose three parameters from HBV as dynamic parameters, $\theta_{d,m}^t$: the shape coefficient of effective flow ($\beta^t$), the shape coefficient evapotranspiration ($\eta^t$), and a dynamic recession coefficient of fast flow ($k_0^t$). The definitions of these variables can be found in Supplementary Table S6. Other parameters ($\theta_{s,m}$) in HBV are assumed to be static in time. An MLP is employed to generate static parameters, and its structure can be found in Supplementary Text S2. Because MERIT basins already have high spatial resolution, only 4 parallel HBV components were used in δHBV2.0, that is, each MERIT basin has 4 subbasin-scale components.

The HBV model is used to predict runoff for MERIT unit basins using dynamic and static parameters along with forcing inputs:

$$q_m^{l:t} = HBV(\theta_{d,m}^{l:t}, \theta_{s,m}, x_m^{l:t}) \tag{6}$$

where $q_m^t$ represents the runoff of MERIT unit basin $m$ at time step $t$.

The runoff of all MERIT unit basins upstream of the target training gage are summed to obtain the total runoff of the gage basin (Eq. 7) and are represented by $q_b^{\prime t}$, which is further routed to the gage basin outlet by an intrinsic unit hydrograph formula (Eq. 8 and 9).

$$q_b^{\prime t} = \sum_{m=1}^{M} q_m^t \tag{7}$$

$$q_b^t = \int_0^t \xi(s) q_b^{\prime s}(t-s) ds \tag{8}$$

$$\xi(s) = \frac{1}{\Gamma(\theta_{ra})\theta_{rb}^{\theta_{ra}}} s^{\theta_{ra}-1} e^{-\frac{1}{\theta_{rb}}} \tag{9}$$

where $M$ is the number of MERIT unit basins within the drainage area of the gage. $\xi(s)$ is the gamma distribution-based unit hydrograph. $\theta_{ra}$ and $\theta_{rb}$ are the static routing parameters that describe the shape of the hydrograph, which is also predicted by the MLP network.

Following the work in Song et al.[50], we employ a loss function based on normalized mean square error:

$$Loss_{NMSE} = \frac{1}{B} \sum_{b=1}^{B} \frac{\sum_{t=1}^{T}(q_b^t - q*_b^t)^2}{(\sigma(q*_b^t) + \epsilon)} \tag{10}$$

where $\sigma(q_b^{*t})$ is the standard deviation of the observed runoff of basin $b$ in the whole training time span, which can avoid overweighting large and/or wet basins in the training. $\epsilon$ is a small value used to avoid a zero denominator.

The multiscale design allows δHBV2 to resolve both spatial heterogeneity and rainfall-runoff nonlinearity in the forcings and attributes at the MERIT-basin resolution, in order to make predictions at that resolution. Such a scale also allows nonlinear and threshold behaviors to better manifest, e.g., concentrated rainfall in a small mountainous region can lead to substantial saturation and runoff but if the same rainfall is spread across a large basin, then very little runoff would occur. δHBV2 simulates runoff for around 2.94 million MERIT unit basins around the world, and then routes the runoff using a differentiable Muskingum-Cunge model (δMC) as described below.

*Differentiable Muskingum-Cunge Routing (δMC)*
Both coarse and fine-resolution models mentioned above inherently use a unit hydrograph formula for channel routing during the training process. Runoff from all MERIT unit basins can be simply averaged and then subsequently routed implicitly to the basin outlet using the unit hydrograph formula. To produce streamflows for every river reach, however, this approach can be cumbersome (we would need a basin delineation at every point of prediction) and inaccurate, especially for major global rivers.

To provide a streamflow product that seamlessly covers all river reaches in the MERIT network, we use an explicit routing model (differentiable Muskingum-Cunge, δMC, developed in Bindas et al.[49]) applied on the MERIT flowlines. The Muskingum-Cunge scheme solves a mass balance (ensuring continuity) and a simplified form of the momentum equation, assuming prismatic shapes for the channels, and conveys the flow from upstream to downstream for every simulated river reach in the network. Muskingum-Cunge has been widely employed for river routing.

Similar to δHBV, δMC incorporates a neural network to learn missing relationships, enabling the improvement of routing results. Here we learn a mapping, represented by a Kolmogorov-Arnold Network (KAN)[80], from attributes of the river reaches, e.g., slope, sinuosity, and upstream catchment area, to the channel geometry and roughness parameters. The channel geometry and roughness, coupled with Manning's formula, allow us to calculate the celerity at each time step, and from this we can calculate parameters for the Muskingum-Cunge scheme[49]. The structure of KAN can be found in Supplementary Text S3. The training objective is to improve flow simulations at a number of downstream gages. Thus the parameters from the trained neural network should be interpreted as effective values to maximize routing accuracy and are not necessarily the ground truth, although future work can incorporate additional datasets and constraints to improve physical realism, as in Chang et al.[81] and Al Mehedi et al.[82].

*Model training and evaluation*
Both versions of the differentiable models were trained from 1980 to 2000 using 4,746 basins (with areas ranging from 21 km$^2$ to 49,821 km$^2$ with a median area of 583 km$^2$, shown in Supplementary Figure S5) and validated from 2001 to 2015 for major river gages that were not directly included in the training dataset (although there could be training gages in their subbasins). The hyperparameters and optimization configurations of the NNs are provided along with their structures in Supplementary Text S4. We first evaluated the performance of δHBV2 over 61 global large rivers (Supplementary Figure S5) from 1981 to 2000 in capturing the water balance over both long-term and seasonal periods, in comparison with previous GWMs stored in The Inter-Sectoral Impact Model Intercomparison Project (phase 2a, ISIMIP2a)[53]. Accurately estimating water quantities is critical for GWMs. We chose this dataset benchmark as it is the most mature available, and we were able to find enough GWMs with uploaded simulations and publications to compare with. Then, we further evaluated our performance over small-to-medium size basins (with areas smaller than 50,000 km$^2$) and another set of untrained stations with ECMWF's operational system, GloFAS[27]. To evaluate the performance of the model, metrics like NSE and KGE are commonly used by the community — detailed information can be found in Supplementary Table

S7. To assess the impacts of past changes (in both precipitation, temperature, and water partitioning), we have also identified estuaries in North America and Europe which are in a global estuary database developed by the Sea Around Us project[83] and have at least one streamflow station upstream. We then evaluate the changes in freshwater inflow through the main stem river into these estuaries.

*Hydrologic Signatures*

Hydrologic signatures are important metrics or indices to describe the statistical and dynamical properties of hydrologic data, e.g. streamflow. In our work, we analyzed models' abilities to capture different hydrologic signatures, e.g. elasticity, slope of the flow duration curve (FDC), and auto-correlation functions (ACFs).

*Elasticity*

Following Zhang et al[62], we suppose that the variability of streamflow at each time interval $i$ is mainly influenced by the variability of precipitation, and thus the streamflow variability can be estimated using the equation:

$$dQ_i = \frac{\delta Q}{\delta P} dP_i \tag{11}$$

and can be further written as:

$$\frac{dQ_i}{Q} = \frac{\delta Q}{\delta P}\frac{P}{Q}\frac{dP_i}{P} = \frac{\delta Q/Q}{\delta P/P}\frac{dP_i}{P} = \varepsilon_{P_i}\frac{dP_i}{P} \tag{12}$$

The above equation assumes that precipitation influences runoff generation at its corresponding time period, which is more appropriate for small-to-medium-size basins and does not account for the significant lag effect observed in large rivers.

In this work, we pre-define the interval for precipitation $i$, and interval for streamflow $j$, and the equation can be written as:

$$\frac{dQ_j}{Q} = \frac{\delta Q}{\delta P}\frac{P}{Q}\frac{dP_i}{P} = \frac{\delta Q/Q}{\delta P/P}\frac{dP_i}{P} = \varepsilon_{P_i}\frac{dP_i}{P} \tag{13}$$

where $P_i$ and $Q_j$ are the summed precipitation and streamflow data at the predefined aggregation intervals $i$ and $j$. For winter elasticity, we consider December, January, and February as the precipitation and streamflow intervals for the Northern Hemisphere and June, July, and August for rivers located in the Southern Hemisphere. As the summer season retains more water compared to winter, we consider the lag effect of precipitation, and define March-August (6 months) as the precipitation interval and June-August as the streamflow response interval for the Northern Hemisphere, and use the opposite months for the Southern Hemisphere. We tested different window lengths, and 6 months gave the highest elasticity for summertime discharge. $P$ and $Q$ in the equation are the mean values for the cumulative precipitation and streamflow over all time intervals. $\varepsilon_{P_i}$ can be interpreted as the sensitivity of streamflow to the variability of precipitation, and can be regarded as the slope of a linear regression line. In this regression, the x-axis represents the precipitation deviations from the mean ($dP_i$), normalized by the mean precipitation ($P$), while the y-axis represents the corresponding calculation for streamflow. To ensure the reliability of the elasticity calculation, we also adopted the F-test and only trusted results with p-values less than 0.1. To better understand the physical meaning of $\varepsilon_{P_i}$, we can assume that if the $\varepsilon_{P_i}$ of one river is equal to 2%, it means that a 1% increase in precipitation would lead to approximately a 2% increase in streamflow.

*Slope of FDC*

Flow duration curves (FDCs) are widely used to characterize streamflow variability. They are constructed by ranking the streamflow data in descending order and calculating the exceedance probability for each value. The slope of an FDC is normally computed over a specific percentile range (e.g. 1% - 33%) to quantify the flow variability. In our work, the slope for daily simulation is defined as:

$$slope = \frac{Log_{10}(Q_{33}) - Log_{10}(Q_1)}{P_{33} - P_1} \tag{14}$$

where $P_{33} - P_1$ indicates the difference between exceedance probabilities at two points on the FDC, and is roughly equal to 0.32 when there is enough data. $P$ represents the exceedance probability and can be calculated as:

$$P_i = \frac{r}{N+1} \tag{15}$$

where $r$ is the rank of streamflow value (starting from 1 for the highest flow) and $N$ is the total number of observations. For the monthly FDC comparison (in the Supplementary Materials Table S3), the log-transformation was not adopted due to the more evenly-distributed monthly time series.

*Auto-correlation Function*

The auto-correlation function (ACF) is normally used to describe the correlation between a time series and its lagged values across multiple timescales, and is expressed as:

$$ACF(k) = \frac{\sum_{t=1}^{n_t-k}(X_t - \bar{X})(X_{t+k} - \bar{X})}{\sum_{t=1}^{n_t-k}(X_t - \bar{X})^2} \tag{16}$$

where $X_t$ is the streamflow value at time $t$, $\bar{X}$ is the mean streamflow value, and $k$ is the lag time.

**Data Availability**

All data used in this work are publicly available. The codes of δHBV1.0 and the multiscale δHBV2 are available at https://doi.org/10.5281/zenodo.14827983. A new repository will be created upon paper acceptance. The estuary datasets were from Global Estuary Database and can be downloaded at https://www.pigma.org/. The MERIT-Basins datasets can be downloaded at https://www.reachhydro.org/home/params/merit-basins. Simulations for other GWMs are available at https://www.isimip.org/. Static geographic attributes were derived from diverse sources and are listed in Supplementary Table S5. MSWEP and MSWX forcing can be downloaded from www.gloh2o.org. The streamflow dataset was compiled and provided by Ather et al.[65], with the GRDC data being described by Ather et al. but downloaded directly from the GRDC at https://grdc.bafg.de/data/data_portal, as they do not permit redistribution of their data.

Supplementary Information for
# Distinct hydrologic response patterns and trends worldwide revealed by physics-embedded learning

**Supplementary Tables**
*Table S1. Mean Annual Runoff (mm/year) and corresponding water export quantities (in the parentheses, units of $10^9$ tons/year) from different GWMs for 33 mixed large rivers*
*Note: In Tables S1 and S2, we only evaluated and compared the simulations with observations when observational data were available.*

| GRDC Station ID | River | Observation | GWM1 | GWM2 | GWM3 | GWM4 | GWM5 | δHBV2 |
|---|---|---|---|---|---|---|---|---|
| 01147010 | Congo | 341.98 (1183.2) | 588.45 (2034.91) | 512.74 (1779.59) | 475.34 (1644.52) | 360.15 (1246.1) | 421.51 (1460.6) | 541.98 (1891.7) |
| 01159100 | Orange | 6.30 (5.56) | 81.11 (70.60) | 13.94 (12.35) | 64.69 (56.71) | 3.02 (2.63) | 57.65 (50.2) | 10.33 (9.24) |
| 01834101 | Niger | 59.79 (121.83) | 135.92 (274.5) | 138.08 (279.29) | 120.54 (243.06) | 65.51 (132.65) | 108.77 (220.68) | 93.41 (187.4) |
| 02906900 | Amur | 201.64 (354.52) | 300.36 (523.87) | 185.66 (323.67) | 180.52 (313.63) | 127.96 (223.89) | 209.70 (366.72) | 139.86 (251.72) |
| 03629000 | Amazon | 1077.37 (5071.9) | 1196.14 (5613.53) | 1097.82 (5172.64) | 1018.52 (4792.44) | 948.73 (4469.4) | 814.25 (3832.7) | 1007.64 (4797.6) |
| 03638050 | Xingu | 575.64 (256.29) | 985.62 (437.04) | 948.09 (419.26) | 913.12 (402.39) | 747.99 (331.46) | 558.43 (244.51) | 729.33 (328.88) |
| 03650481 | Parnaiba | 70.84 (22.87) | 308.64 (99.63) | 222.97 (71.98) | 269.47 (86.99) | 107.54 (34.72) | 124.53 (40.2) | 107.64 (34.75) |
| 03651805 | Sao Francisco | 319.61 (64.41) | 586.05 (118.6) | 492.73 (99.77) | 451.59 (91.02) | 361.16 (73.74) | 503.76 (102.26) | 355.90 (73.65) |
| 03667060 | Paraguai | 196.79 (92.42) | 493.57 (231.9) | 364.25 (170.21) | 360.93 (169.43) | 295.46 (137.98) | 421.17 (197.83) | 321.07 (154.35) |
| 04115200 | Columbia | 269.37 (166.37) | 331.65 (203.61) | 242.66 (149.52) | 218.42 (134.34) | 134.32 (83.19) | 181.04 (111.49) | 258.86 (161.52) |
| 04127800 | Mississippi | 194.35 (581.09) | 356.37 (1065.09) | 191.13 (572.71) | 254.41 (760.79) | 166.44 (493.78) | 279.03 (835.49) | 196.47 (600.24) |
| 04143550 | Saint Lawrence | 311.59 (240.77) | 612.44 (474.93) | 337.56 (261.7) | 240.72 (186.92) | 510.75 (395.41) | 270.89 (209.94) | 402.02 (319.86) |
| 04152450 | Colorado | 48.85 (14.15) | 121.35 (35.12) | 63.30 (18.07) | 102.90 (29.23) | 11.18 (3.19) | 78.35 (22.63) | 80.40 (23.5) |
| 04207900 | Fraser | 393.13 (85.77) | 468.82 (102.0) | 398.26 (86.95) | 275.62 (59.91) | 253.20 (55.55) | 296.79 (64.61) | 398.28 (87.71) |
| 04208150 | Mackenzie | 150.40 (240.7) | 231.80 (367.84) | 195.70 (313.21) | 107.46 (174.26) | 114.59 (183.86) | 127.45 (203.91) | 133.67 (216.56) |
| 04213550 | Saskatchewan | 48.33 (16.82) | 182.79 (63.47) | 104.22 (36.08) | 136.72 (47.19) | 54.73 (18.95) | 95.40 (33.05) | 66.20 (23.39) |
| 04213650 | Assiniboine | 8.60 (1.35) | 144.58 (22.32) | 68.77 (10.61) | 155.02 (23.77) | 21.48 (3.34) | 47.89 (7.47) | 19.89 (3.14) |
| 04213680 | Red River of The North | 45.95 (4.94) | 194.76 (20.57) | 80.91 (8.54) | 161.92 (16.96) | 47.37 (5.05) | 88.62 (9.4) | 42.48 (4.59) |
| 04213800 | Winnipeg | 221.68 (28.33) | 445.45 (56.67) | 209.91 (26.58) | 177.55 (22.63) | 266.59 (33.95) | 183.16 (23.43) | 214.63 (27.83) |
| 04214260 | Churchill | 86.48 (19.73) | 317.32 (73.22) | 213.40 (49.5) | 104.29 (24.14) | 149.12 (34.3) | 1.93 (0.45) | 111.18 (26.21) |
| 04214520 | Albany | 232.30 (27.68) | 534.76 (63.36) | 382.26 (45.12) | 301.08 (35.76) | 366.60 (43.25) | 220.44 (26.18) | 288.37 (35.04) |
| 04355100 | Grande De Santiago | 38.75 (5.0) | 232.71 (29.88) | 126.43 (16.17) | 151.38 (19.39) | 13.95 (1.78) | 221.92 (28.51) | 57.03 (7.44) |
| 05101200 | Burdekin | 56.86 (7.5) | 212.50 (27.62) | 81.49 (10.85) | 165.81 (21.75) | 39.97 (5.11) | 63.38 (8.31) | 83.02 (11.04) |
| 05101301 | Fitzroy | 30.97 (4.36) | 217.47 (29.75) | 50.78 (7.06) | 170.33 (23.54) | 24.37 (3.18) | 84.86 (11.69) | 57.62 (8.08) |
| 05204250 | Darling | 4.40 (2.35) | 194.85 (100.17 | 31.29 (16.32) | 115.18 (59.62) | 14.70 (7.72) | 75.36 (38.91) | 53.86 (27.63) |
| 05410100 | Cooper | 5.08 (1.24) | 170.73 (40.16) | 18.97 (4.57) | 96.75 (22.98) | 19.59 (4.47) | 24.39 (5.78) | 99.77 (23.7) |
| 06340110 | Elbe | 167.56 (21.64) | 378.97 (49.71) | 252.82 (33.16) | 272.99 (35.91) | 193.31 (25.2) | 418.91 (54.97) | 188.74 (25.26) |
| 06435060 | Rhine | 464.82 (74.44) | 724.87 (116.37) | 570.48 (91.58) | 572.19 (91.87) | 556.97 (89.08) | 654.42 (104.92) | 500.00 (81.7) |
| 06442600 | Danube | 352.47 (73.19) | 569.48 (119.03) | 398.80 (83.11) | 405.38 (84.78) | 370.39 (77.19) | 537.88 (112.33) | 403.51 (85.38) |
| 06955430 | Neva | 283.41 (80.37) | 525.45 (149.22) | 446.98 (126.87) | 316.99 (89.9) | 342.68 (97.24) | 134.17 (38.45) | 323.60 (94.02) |

| 06970250 | Severnaya Dvina | 301.46 (105.72) | 441.36 (154.37) | 421.58 (147.91) | 299.40 (104.7) | 324.29 (113.29) | 104.91 (37.0) | 288.67 (103.65) |
|---|---|---|---|---|---|---|---|---|
| 06977100 | Volga | 190.92 (260.18) | 357.10 (484.78) | 287.70 (392.23) | 289.90 (393.68) | 204.08 (276.0) | 183.42 (248.62) | 201.27 (280.53) |
| 06978250 | Don | 58.18 (22.02) | 230.92 (86.58) | 173.85 (65.69) | 228.47 (85.49) | 59.42 (22.09) | 258.63 (96.41) | 56.53 (21.78) |
| Summed value | | (9258.81) | (13240.25) | (10902.87) | (10085.54) | (8728.77) | (8849.67) | (10029.2) |
| Bias | | | (3981.44) | (1644.06) | (826.73) | (-530.04) | (-409.14) | (770.39) |

*Table S2.* Mean Annual Runoff (mm/year) and corresponding water quantities (units of $10^9$ tons/year) from different GWMs for 28 natural basins.

| GRDC Station ID | River | Observation | GWM1 | GWM2 | GWM3 | GWM4 | GWM5 | δHBV2 |
|---|---|---|---|---|---|---|---|---|
| 01159103 | Orange | 5.31 (4.57) | 79.43 (68.31) | 13.83 (11.89) | 61.10 (52.54) | 2.98 (2.56) | 57.00 (49.02) | 10.19 (8.76) |
| 02469260 | Mekong | 531.23 (289.52) | 736.68 (401.49) | 669.32 (364.78) | 617.81 (336.71) | 424.69 (231.46) | 555.21 (302.59) | 449.12 (244.77) |
| 02646200 | Ganges | 361.45 (305.9) | 646.40 (547.05) | 562.43 (475.98) | 593.97 (502.68) | 389.63 (329.75) | 519.95 (440.04) | 342.29 (289.68) |
| 02903429 | Lena | 257.78 (231.23) | 213.51 (191.52) | 217.42 (195.02) | 160.17 (143.67) | 151.86 (143.67) | 231.54 (207.69) | 325.60 (292.06) |
| 02903600 | Aldan | 241.74 (168.25) | 230.36 (160.33) | 220.08 (153.17) | 183.67 (127.83) | 178.78 (124.43) | 260.32 (181.19) | 380.95 (265.14) |
| 02906700 | Amur | 163.96 (267.25) | 300.79 (490.28) | 192.24 (313.35) | 180.86 (294.8) | 124.90 (203.59) | 216.03 (352.12) | 113.26 (184.62) |
| 02906901 | Amur | 202.32 (362.16) | 310.91 (556.53) | 193.04 (345.54) | 190.34 (340.71) | 141.52 (253.32) | 222.40 (398.1) | 155.62 (278.56) |
| 02909150 | Yenisey | 246.23 (600.76) | 233.24 (569.11) | 248.25 (605.74) | 190.56 (464.97) | 179.83 (438.79) | 200.16 (488.4) | 237.64 (579.85) |
| 02909152 | Yenisey | 192.12 (338.12) | 207.05 (364.41) | 211.55 (372.33) | 148.59 (261.52) | 127.44 (224.3) | 165.37 (291.05) | 181.92 (320.17) |
| 02909153 | Yenisey | 169.27 (236.97) | 175.47 (245.65) | 184.29 (258.0) | 118.62 (166.07) | 88.68 (124.15) | 163.28 (228.6) | 161.87 (226.62) |
| 02911097 | Irtysh | 42.78 (64.17) | 102.67 (154.01) | 83.99 (125.99) | 109.11 (163.67) | 29.72 (44.58) | 100.00 (150.0) | 45.77 (68.66) |
| 02911100 | Irtysh | 30.27 (23.28) | 49.43 (38.01) | 38.11 (29.31) | 59.73 (45.93) | 9.68 (7.45) | 88.71 (68.22) | 23.98 (18.44) |

| ID | River | | | | | | | |
|---|---|---|---|---|---|---|---|---|
| 02912600 | Ob' | 136.64 (403.07) | 218.58 (644.81) | 201.44 (594.25) | 185.64 (547.65) | 130.05 (383.66) | 152.06 (448.56) | 135.63 (400.11) |
| 02912602 | Ob' | 111.49 (299.9) | 195.99 (527.21) | 180.21 (484.76) | 172.01 (462.71) | 105.58 (284.0) | 135.89 (365.54) | 110.35 (296.85) |
| 03206720 | Orinoco | 1233.23 (1030.98) | 1321.96 (1105.16) | 1279.73 (1069.85) | 1130.67 (945.24) | 1080.01 (902.89) | 1295.09 (1082.69) | 1323.00 (1106.03) |
| 03264500 | Parana | 489.16 (476.93) | 772.74 (753.42) | 552.16 (538.36) | 527.36 (514.18) | 520.36 (507.35) | 683.87 (666.77) | 494.93 (482.55) |
| 03617110 | Mamore | 441.60 (268.9) | 741.62 (451.64) | 606.75 (369.51) | 597.69 (363.99) | 409.48 (249.37) | 491.27 (299.18) | 411.51 (250.61) |
| 03635035 | Madera | 654.70 (713.54) | 912.98 (995.15) | 802.49 (874.72) | 757.92 (826.13) | 581.62 (633.97) | 620.17 (675.99) | 592.30 (645.61) |
| 03635040 | Madera | 650.55 (634.87) | 896.90 (875.38) | 772.05 (753.52) | 734.15 (716.53) | 556.07 (542.73) | 624.49 (609.5) | 568.95 (555.29) |
| 03649900 | Tocantins | 448.30 (327.76) | 797.09 (582.89) | 696.52 (509.35) | 640.15 (468.12) | 556.66 (407.07) | 609.44 (445.67) | 490.31 (358.55) |
| 03649901 | Tocantins | 474.78 (327.97) | 890.58 (615.32) | 773.65 (534.53) | 712.92 (492.57) | 627.15 (433.31) | 681.33 (470.74) | 519.47 (358.91) |
| 03649950 | Tocantins | 477.98 (354.77) | 863.47 (640.95) | 751.41 (557.77) | 693.91 (515.09) | 609.81 (452.66) | 650.17 (482.62) | 537.84 (399.24) |
| 04103200 | Yukon River | 246.26 (204.74) | 186.01 (154.64) | 225.39 (187.39) | 164.40 (136.68) | 154.63 (128.56) | 134.91 (112.16) | 178.66 (148.54) |
| 04103550 | Yukon | 210.52 (107.03) | 163.39 (83.07) | 210.83 (107.19) | 149.50 (76.01) | 133.79 (68.02) | 121.25 (61.65) | 166.40 (84.6) |
| 04115201 | Columbia | 312.52 (207.9) | 388.35 (258.4) | 292.35 (194.52) | 268.51 (178.66) | 182.93 (121.72) | 266.74 (177.48) | 325.41 (216.52) |
| 04122900 | Missouri | 66.61 (90.43) | 185.53 (251.89) | 78.57 (106.68) | 140.09 (190.2) | 48.81 (66.26) | 137.76 (187.03) | 69.66 (94.57) |
| 04122901 | Missouri | 53.33 (69.16) | 166.29 (215.65) | 67.08 (86.99) | 126.46 (163.99) | 38.34 (49.72) | 127.86 (165.81) | 58.45 (75.79) |
| 04123050 | Ohio | 495.26 (260.38) | 685.27 (360.29) | 417.41 (219.46) | 473.44 (248.92) | 376.47 (197.94) | 454.16 (238.78) | 468.91 (246.54) |
| 04127503 | Mississippi | 112.64 (203.33) | 242.57 (437.9) | 118.65 (214.2) | 175.51 (316.83) | 103.11 (186.15) | 202.00 (364.65) | 114.40 (206.52) |
| 06542510 | Danube | 301.23 (158.38) | 508.70 (267.48) | 334.64 (175.96) | 369.78 (194.44) | 283.42 (149.03) | 442.54 (232.69) | 389.66 (204.89) |
| Summed value | | (8633.74) | (12656.10) | (10481.92) | (9850.86) | (7624.36) | (9855.65) | (8351.85) |
| Bias | | | (4022.36) | (1848.18) | (1217.12) | (-1009.38) | (1221.91) | (-281.89) |

*Table S3.* Spatial $R^2$ values of different hydrologic signatures at the monthly scale. The monthly hydrologic signatures were calculated for each basin, and their $R^2$ values were computed spatially across 28 natural rivers. Note that the slope of FDC was calculated based on the 1% and 33% exceedance probabilities, and only stations with at least 80% data availability were considered reliable for this calculation. The lag in parentheses after ACF indicates the number of lagged months used to calculate the autocorrelation function. The following metrics were calculated from 1981-2000, except for elasticity which was calculated using data from 1981-2010 to ensure sufficient data for statistical analysis.

| Models | Slope of FDC | Summer elasticity | Winter elasticity | ACF (1) | ACF (2) |
|---|---|---|---|---|---|
| GWM0 | -0.18 | 0.36 | 0.76 | 0.57 | 0.44 |
| GWM1 | -0.75 | -0.02 | 0.72 | 0.93 | 0.16 |
| GWM2 | -2.48 | 0.32 | 0.72 | 0.53 | 0.28 |
| GWM3 | -1.40 | 0.38 | -1.36 | -0.84 | -0.28 |
| GWM4 | -0.11 | -1.06 | -11.03 | 0.11 | -0.26 |
| GWM5 | -0.55 | 0.01 | 0.67 | -0.53 | -0.32 |
| δHBV2 | 0.55 | 0.76 | 0.79 | 0.52 | 0.71 |

*Table S4.* Median performance metrics at daily scale for LSTM, δHBV1.0, and δHBV2.0 across different continents, based on 4,746 training basins smaller than 50,000 km². Bold font indicates the best value in each category.

| Region | Model | NSE | KGE | Bias (mm/day) | RMSE (mm/day) | FLV | FHV | Low RMSE (mm/day) | High RMSE (mm/day) |
|---|---|---|---|---|---|---|---|---|---|
| Global | LSTM | **0.732** | 0.734 | -0.018 | 0.637 | 27.035 | -12.854 | **0.063** | 1.541 |
|  | δHBV1 | 0.657 | 0.662 | -0.025 | 0.732 | 25.054 | -15.225 | 0.066 | 1.830 |
|  | δHBV2 | 0.727 | **0.742** | 0.02 | 0.657 | 37.731 | **-8.251** | 0.068 | **1.442** |
| Europe | LSTM | **0.682** | **0.699** | -0.042 | 0.499 | 18.052 | -17.360 | **0.065** | 1.160 |
|  | δHBV1 | 0.574 | 0.610 | -0.042 | 0.545 | 15.076 | -20.135 | 0.069 | 1.447 |
|  | δHBV2 | 0.648 | 0.696 | 0.02 | 0.506 | 36.537 | **-13.547** | 0.085 | **1.105** |

| Region | Model | | | | | | | | |
|---|---|---|---|---|---|---|---|---|---|
| North America | LSTM | 0.783 | 0.787 | -0.016 | 0.786 | 25.960 | -11.007 | 0.056 | 1.781 |
| | δHBV1 | 0.718 | 0.727 | -0.04 | 0.894 | 22.334 | -15.108 | 0.05 | 2.159 |
| | δHBV2 | **0.788** | **0.799** | 0.003 | 0.778 | 29.27 | **-6.817** | **0.048** | **1.598** |
| Africa | LSTM | 0.438 | **0.381** | -0.032 | 0.467 | 219.847 | -36.081 | 0.046 | 1.388 |
| | δHBV1 | 0.380 | 0.249 | -0.052 | 0.503 | 104.963 | -49.526 | 0.03 | 1.922 |
| | δHBV2 | **0.453** | 0.362 | -0.044 | 0.481 | 59.989 | -41.013 | 0.016 | 1.564 |
| South Asia | LSTM | **0.745** | **0.783** | -0.118 | 2.140 | 8.751 | -7.821 | 0.234 | 4.628 |
| | δHBV1 | 0.595 | 0.610 | -0.478 | 2.743 | -30.628 | -19.002 | 0.520 | 7.244 |
| | δHBV2 | 0.711 | 0.755 | 0.054 | 2.246 | 17.661 | -6.85 | 0.269 | 4.830 |
| Oceania and South Asian Islands | LSTM | 0.700 | **0.661** | 0.015 | 0.465 | 160.734 | -1.705 | 0.032 | 1.600 |
| | δHBV1 | 0.620 | 0.484 | 0.064 | 0.562 | 449.635 | -2.209 | 0.062 | 1.847 |
| | δHBV2 | 0.700 | 0.630 | 0.025 | 0.508 | 94.063 | 2.291 | 0.021 | 1.684 |
| South America | LSTM | **0.658** | **0.681** | 0.067 | 0.686 | 41.938 | -11.342 | 0.138 | 1.616 |
| | δHBV1 | 0.551 | 0.586 | 0.194 | 0.829 | 33.523 | -3.497 | 0.120 | 1.564 |
| | δHBV2 | 0.597 | 0.638 | 0.208 | 0.755 | 50.423 | -2.063 | 0.171 | 1.448 |

*Table S5.* *Summary of forcings and attributes used in model training and simulation (and the details of each attribute can be found in Song et al.[1])*

| Variables | Description | Units |
|---|---|---|
| P | Daily precipitation from MSWEP | mm/day |

| Variable | Description | Unit |
|---|---|---|
| Temp | Daily mean temperature from MSWX | °C |
| PET | Potential daily evapotranspiration obtained using Hargreaves method based on maximal and minimal temperature from MSWX | mm/day |
| catchment area/upstream area | Upstream area of a MERIT unit basin is used in δHBV2.0 and the gage basin area is used in δHBV1.0 | $km^2$ |
| ETPOT_Hargr | Potential evapotranspiration obtained using the Hargreaves method | mm/year |
| FW | Fraction of open water data in the Global Lakes and Wetlands Database | 1 |
| HWSD_clay | Proportion of clay in soil in the Harmonized World Soil Database (HWSD 1.2 and 2.0) | % |
| HWSD_gravel | Proportion of gravel in soil in the Harmonized World Soil Database (HWSD 1.2 and 2.0) | % |
| HWSD_sand | Proportion of sand in soil in the Harmonized World Soil Database (HWSD 1.2 and 2.0) | % |
| HWSD_silt | Proportion of silt in soil in the Harmonized World Soil Database (HWSD 1.2 and 2.0) | % |
| NDVI | The Normalized Difference Vegetation Index | - |
| Porosity | Porosity from the Global Hydrogeology Maps (GLHYMPS) | - |
| SoilGrids_clay | Proportion of clay in the SoilGrids dataset | % |
| SoilGrids_sand | Proportion of clay in the SoilGrids dataset | % |
| SoilGrids_silt | Proportion of clay in the SoilGrids dataset | % |
| aridity | Aridity calculated from forcings | - |
| glaciers | Glacier fraction from the Global Lithological Map (GLiM) | 1 |
| meanP | Mean precipitation from WorldClim | mm/year |
| meanTa | Mean temperature from WorldClim | °C |
| Mean elevation | Mean elevation | m |
| meanslope | Mean slope | degree |
| permafrost | Permafrost | 1 |
| permeability | Permeability from the Global Hydrogeology Maps (GLHYMPS) | $\log(m^2)$ |
| seasonality_P | Precipitation seasonality index calculated using the Seasonality Index (SI) method | - |
| seasonality_PET | Potential Evapotranspiration seasonality index calculated using the Seasonality Index (SI) method | - |
| snow_fraction | Snow fraction | 1 |
| snowfall_fraction | Snowfall fraction | 1 |

***Table S6.*** *HBV module details, including their equations, fluxes, and parameters.*

| Modules | Equations | Fluxes | Parameters |
|---|---|---|---|
| Snow accumulation and melt | $\frac{dS_p}{dt} = P_s + R_{fz} - S_{melt}$ <br> $P_s = P$ if $T < \theta_{TT}$, otherwise 0 <br> $R_{fz} = (\theta_{TT} - T)\theta_{DD}\theta_{rfz}$ <br> $s_{melt} = (T - \theta_{TT})\theta_{DD}$ <br> $\frac{dS_{liq}}{dt} = s_{melt} - R_{fz} - I_{snow}$ <br> $I_{snow} = s_{liq} - \theta_{CWH}S_p$ | $S_p$: storage in snow <br> $P_s$: precipitation as snow <br> $R_{fz}$: refreezing of liquid snow <br> $S_{melt}$: snowmelt as water equivalent <br> $S_{liq}$: liquid water content in the snowpack <br> $I_{snow}$: snowmelt infiltration to soil moisture | $\theta_{TT}$: threshold temperature for snowfall [°C], range: [-2.5,2.5]. <br> $\theta_{DD}$: degree-day factor [mm°C$^{-1}$day$^{-1}$], range: [0,0.1]. <br> $\theta_{rfz}$: refreezing coefficient [-], range: [0.5,10]. <br> $\theta_{CWH}$: water holding capacity as a fraction of the current snowpack [-], range: [0,0.2]. |
| Soil moisture and evapotranspiration | $\frac{dS_S}{dt} = I_{snow} + P_r - P_{eff} - E_x - E_T(+C_r)$ <br><br> $P_r = P$ if $T > \theta_{TT}$ otherwise 0 <br> $P_{eff} = \min\left(\left(\frac{S_S}{\theta_{FC}}\right)^\beta, 1\right)(P_r + I_{snow})$ <br> $E_x = (S_S - \theta_{FC})/dt$ <br> $E_T = \min\left(\left(\frac{S_S}{\theta_{FC}\theta_{LP}}\right)^\gamma, 1\right)E_P$ <br> $C_r = \theta_C * S_{LZ} * (1 - \frac{S_S}{\theta_{FC}})$ | $S_S$: storage in soil moisture <br> $S_{LZ}$: current storage in the lower subsurface zone <br> $P_r$: precipitation as rain <br> $P_{eff}$: effective flow to the upper subsurface zone <br> $E_x$: rainfall excess <br> $E_T$: actual evapotranspiration. <br> $C_r$: an upward flow from the lower subsurface zone. This flux is only used in δHBV1.1p | $\theta_C$: time parameter [day$^{-1}$], range: [0,1]. <br> $\theta_{FC}$: maximum soil moisture (field capacity) [mm], range: [50,1000]. <br> $\theta_{LP}$: vegetation wilting point [-], range: [0.2,1]. <br> $\beta$: a parameter influencing the shape of the soil moisture function [-], range: [1,6]. <br> $\gamma$: a parameter influencing the shape of the evapotranspiration function [-], range: [0.3,5]. |
| Runoff generation | $\frac{dS_{UZ}}{dt} = P_{eff} + E_x - perc - Q_0 - Q_1$ <br> $perc = \min(\theta_{perc}, S_{UZ}/dt)$ <br> $Q_0 = \theta_{K_0}(S_{UZ} - \theta_{UZL})$ <br> $Q_1 = \theta_{K_1}S_{UZ}$ <br> $\frac{dS_{LZ}}{dt} = perc - Q_2(-C_r)$ <br> $Q_2 = \theta_{K_0}S_{LZ}$ <br> $Q' = Q_0 + Q_1 + Q_2$ <br> $Q_b(t) = \int_0^t \xi(s)Q'(t-s)ds$ <br> $\xi(s) = \frac{1}{\Gamma(\theta_a)\theta_b^{\theta_a}} s^{\theta_a - 1} e^{-\frac{1}{\theta_b}}$ | $S_{UZ}$: storage in the upper subsurface zone <br> $perc$: percolation to the lower subsurface zone <br> $Q_0$: near surface flow <br> $Q_1$: interflow <br> $Q_2$: baseflow <br> $S_{LZ}$: storage in the lower subsurface zone <br> $Q_b$: simulated streamflow for basin $b$ | $\theta_{perc}$: percolation flow rate [mm/day], range: [0,10]. <br> $\theta_{K_0}$ (range: [0,10]), $\theta_{K_1}$ (range: [0.05,0.9]), and $\theta_{K_2}$ (range: [0.001,0.2]): recession coefficients [day$^{-1}$] <br> $\theta_a$ (range: [0,2.9]) and $\theta_b$ (range: [0,6.5]): two routing parameters [-] <br> $\theta_{UZL}$: maximum storage of the upper soil layer [mm], range: [0,100]. |

***Table S7.*** *Model evaluation metrics*

| Metric | Equation | Description |
|---|---|---|
| NSE | $NSE = 1 - \frac{\sum_{i=1}^{i=N}(Q_i^{obs} - Q_i^{sim})^2}{\sum_{i=1}^{i=N}(Q_i^{obs} - \overline{Q_i^{obs}})^2}$ | Ratio of the model error variance to the observed variance, where 1 indicates a perfect model and 0 indicates performance equivalent to predicting with the long-term mean |
| KGE | $KGE = 1 - \sqrt{(r-1)^2 + (a-1)^2 + (b-1)^2}$ | A comprehensive assessment of Pearson's correlation ($r$), variability ($a$), and bias ($b$). |
| Bias | $Bias = \frac{1}{n}\sum_{i=1}^{n}(Q_i^{sim} - Q_i^{obs})$ | Average difference between simulated and observed values |
| RMSE | $$RMSE = \sqrt{\frac{\sum_{i=1}^{i=N}(Q_i^{obs} - Q_i^{sim})^2}{n}}$$ | Root-mean-square error (RMSE) of runoff |
| highRMSE | $$RMSE = \sqrt{\frac{\sum_{i=1}^{i=N}(Q_i^{obs} - Q_i^{sim})^2}{n}}$$ | RMSE of the top 2% of the streamflow range |
| lowRMSE | $$RMSE = \sqrt{\frac{\sum_{i=1}^{i=N}(Q_i^{obs} - Q_i^{sim})^2}{n}}$$ | RMSE of the bottom 30% of the streamflow range |
| FLV | $$FLV = \frac{\sum_{i=1}^{n}(Q_i^{sim\_low} - Q_i^{obs\_low})}{n \sum_{i=1}^{n}(Q_i^{obs\_low})} * 100\%$$ | Percentile bias of the bottom 30% low flow range |
| FHV | $$FHV = \frac{\sum_{i=1}^{n}(Q_i^{sim\_high} - Q_i^{obs\_high})}{\sum_{i=1}^{n}(Q_i^{obs\_high})} * 100\%$$ | Percentile bias of the top 2% peak flow range |

## Supplementary Figures

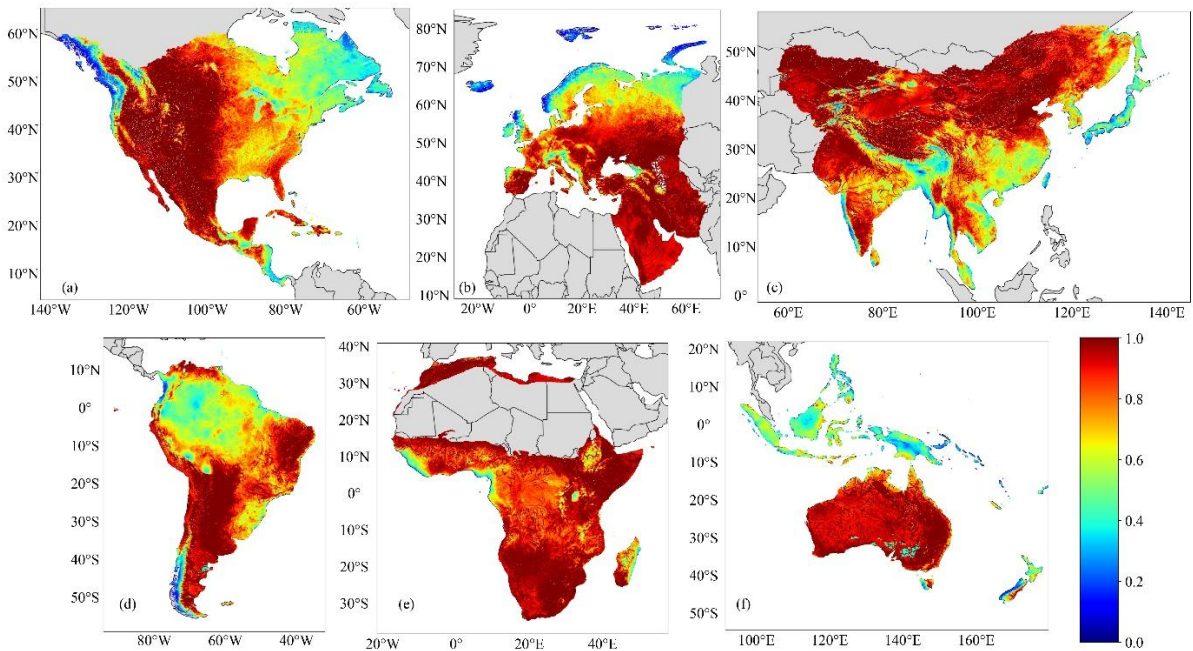

***Figure S1.*** *Distribution of 2001-2020 average ET/P of MERIT unit basins in different continents. (a) North America; (b) Europe; (c) South Asia; (d) South America; (e) Africa without Sahara Desert; (f) Oceania and South Asian Islands.*

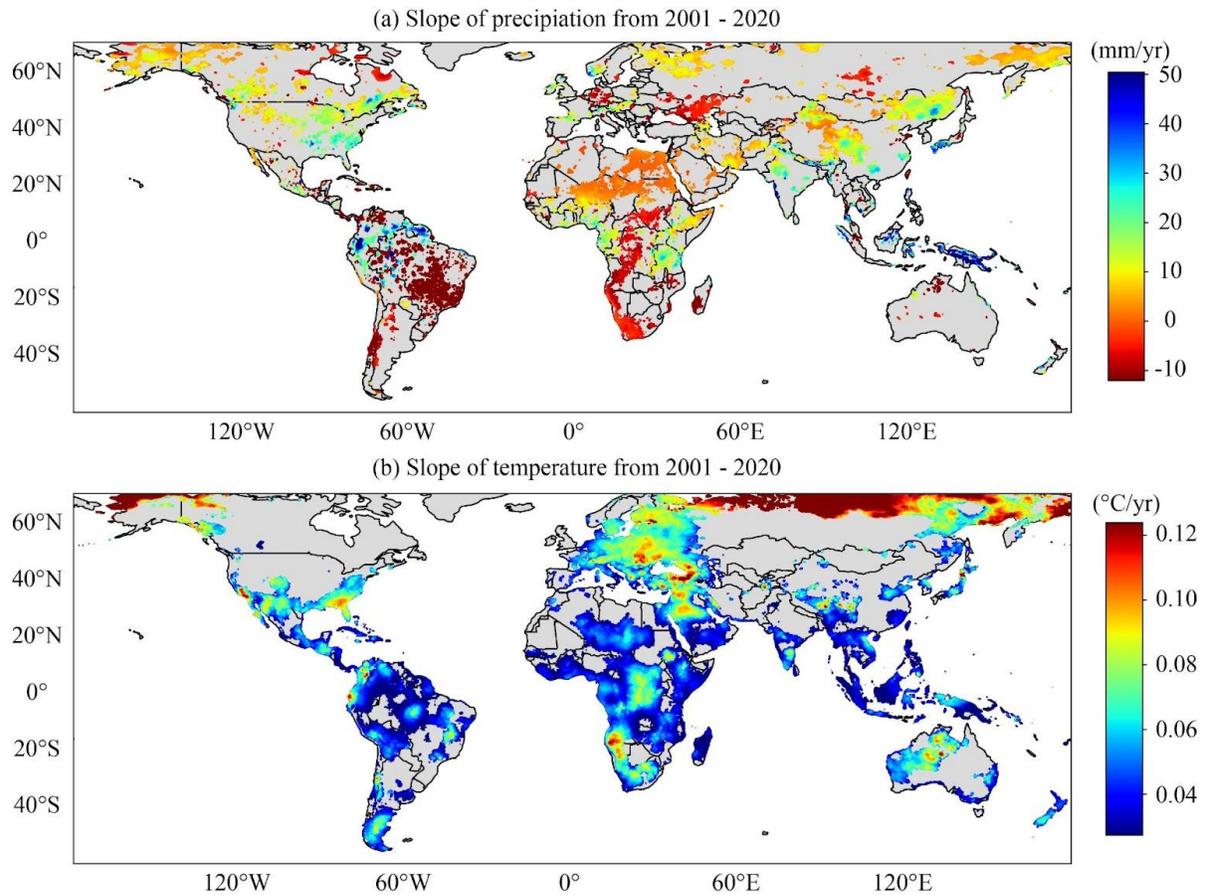

***Figure S2.*** *(a) Decadal-scale trends of precipitation (mm/year) and (b) temperature (°C /year). Only MERIT unit basins that have statistically significant trends in 2001-2020 in Mann-Kendall test ($p < 0.05$) are plotted.*

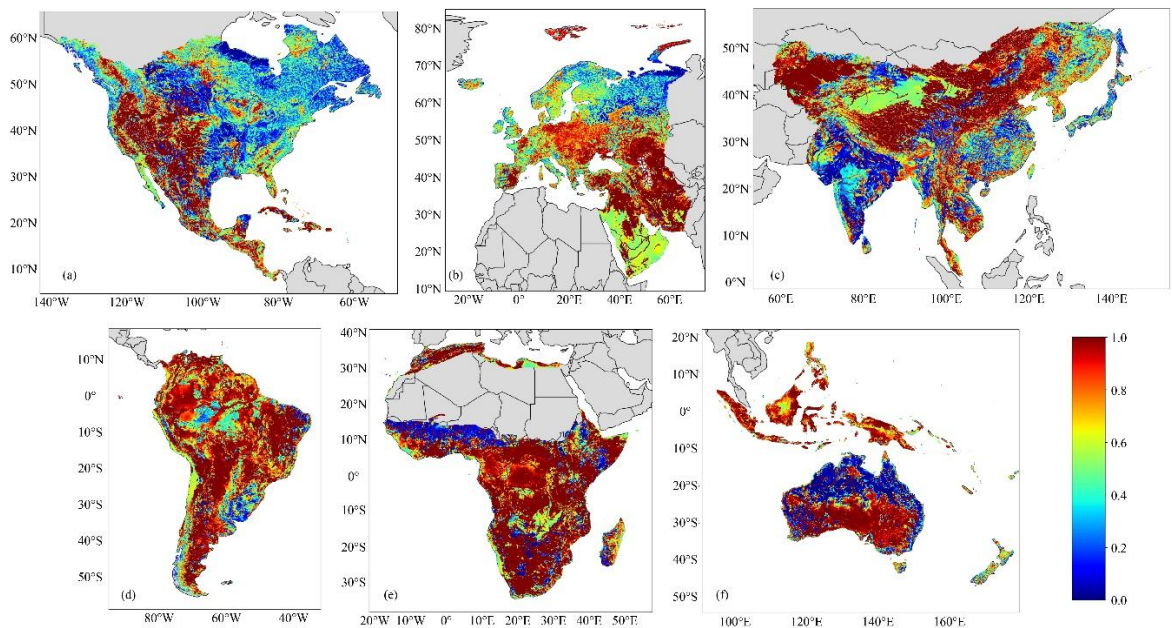

*Figure S3.* Distribution of 2001-2020 average baseflow index of MERIT unit basins in different continents. (a) North America; (b) Europe; (c) South Asia; (d) South America; (e) Africa without Sahara Desert; (f) Oceania and South Asian Islands.

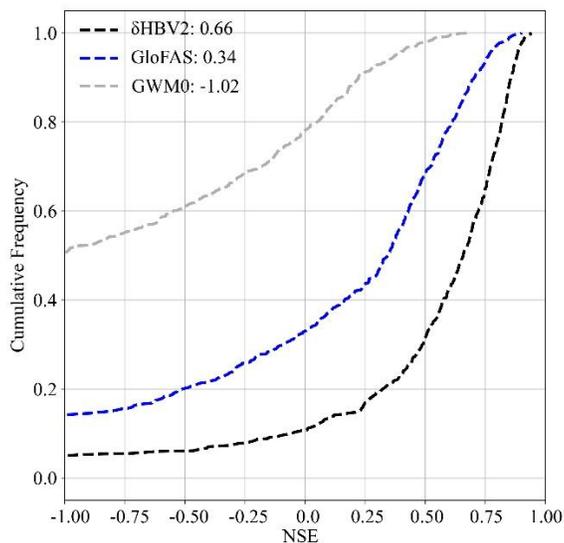

*Figure S4.* CDF plot of Nash-Sutcliffe Efficiency for 664 stations from 1980-01-01 to 2010-12-31, comparing δHBV2, GloFAS, and GWM0. All basins with an area smaller than 50,000km$^2$, and the area difference between model and observation less than 20% were selected.

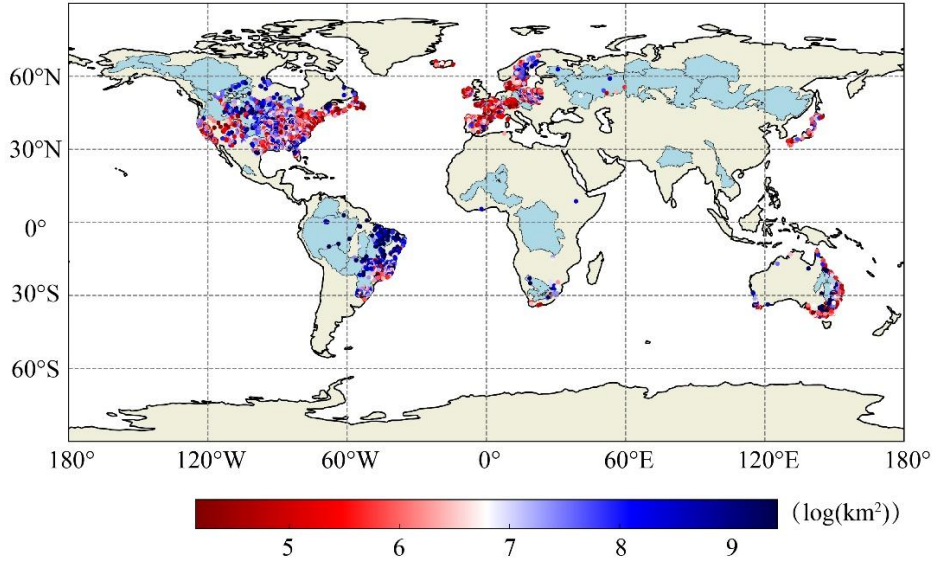

***Figure S5.*** *Locations of 4,746 global training stations. The color of stations indicates their log-transformed catchment size. The 61 large river basins used for further evaluation are marked as light blue polygons.*

**Supplementary Text**

**Text S1: Long Short-Term Memory**
LSTM for both streamflow prediction and dynamic parameterization in δHBV share the same structure, including a linear input layer, input, forget, and output gates to selectively add, retain, or discard information, and cell states to maintain short-term and long-term dependencies (Eq. S1 to Eq. S8).

| | | |
|---|---|---|
| Input transformation: | $x^t = \text{ReLU}(W_I I^t + b_I)$ | (S1) |
| Input gate: | $i^t = \sigma(\mathcal{D}(W_{ix} x^t) + \mathcal{D}(W_{ih} h^{t-1}) + b_i)$ | (S2) |
| Forget gate: | $f^t = \sigma(\mathcal{D}(W_{fx} x^t) + \mathcal{D}(W_{fh} h^{t-1}) + b_f)$ | (S3) |
| Input node: | $g^t = \tanh(\mathcal{D}(W_{gx} x^t) + \mathcal{D}(W_{gh} h^{t-1}) + b_g)$ | (S4) |
| Output gate: | $o^t = \sigma(\mathcal{D}(W_{ox} x^t) + \mathcal{D}(W_{oh} h^{t-1}) + b_o)$ | (S5) |
| Cell state: | $s^t = g^t \odot i^t + s^{t-1} \odot f^t$ | (S6) |
| Hidden state: | $h^t = \tanh(s^t) \odot o^t$ | (S7) |
| Output: | $y^t = W_{hy} h^t + b_y$ | (S8) |

where $I^t$ is the concatenation of the normalized meteorological forcings and static attributes. $\mathcal{D}$ is the dropout operator, which randomly sets the weights of some neurons to zero to prevent overfitting. Variables $i$, $f$, and $o$ are the input, forget, and output gates of LSTM, respectively. $g$ is the input node. $s$ and $h$ represent the hidden states and cell states, respectively. **ReLU, tanh** and $\sigma$ represent the Rectified Linear Unit, the hyperbolic tangent, and sigmoid activation functions, respectively. $W$ and $b$ are the weights of neurons and bias of each layer, respectively. $\odot$ is element-wise multiplication. $y$ is the predicted variable. The predicted variables of the LSTM in δHBV are the HBV parameters, which are further transformed to a range of 0 to 1 using a sigmoid activation function and then rescaled to their actual values based on the parameter ranges provided in Table S6.

**Text S2: Multilayer Perceptron**
The MLP for static parameterization in δHBV2.0 has the following structure:

$$\begin{aligned} \text{Input layer:} \quad & x_1^t = \mathcal{D}(ReLU(W_I I^t + b_I)) \\ \text{Hidden layers:} \quad & x_i^t = \mathcal{D}(ReLU(W_h x_{i-1}^t + b_h)) \\ \text{Hidden layer:} \quad & y^t = \sigma(ReLU(W_o x_i^t + b_o)) \end{aligned} \quad (S9)$$

where $I$ is the normalized static attributes. $W$ and $b$ with different subscripts are the weights and bias of each layer. There are 6 hidden layers, and $x_{i-1}^t$ in each hidden layer represents the outputs from the previous layer. LSTM and MLP are trained together in δHBV2.0. Similarly, static and routing parameters predicted by MLP are rescaled to the real parameter values with the parameter ranges provided in Table S6.

**Text S3: Kolmogorov-Arnold Networks**
The KANs to predict the static Muskingum Cunge parameters in δMC has the following structure:

$$\begin{aligned} \text{Continuous function:} \quad & \phi_{ij}(x_i) = \alpha_{ij} SiLu(x_i) + \beta_{ij} \sum_i c_i B_i(x_i) \\ \text{KAN Layer:} \quad & K_j = \sum_{i=1}^{n} \phi_{ij}(x_i) \\ \text{Output layer:} \quad & y = \sigma(K) \end{aligned} \quad (S10)$$

where $x_i$ is the input of node $i$ in the input layer; subscript $j$ denotes the node $j$ in the output layer. $\sum_i c_i B_i(x_i)$ is the weighted sum of basic functions of all input nodes, where $B$ is the B-spline function and $c_i$ is the weight assigned to node $i$. The scaling parameter $\beta_{ij}$ is applied to node $i$ to compute output $K_j$ of the node $j$ in the KAN layer. The $SiLu$ activation function, defined as $SiLu(x) = x/(1 + e^{-z})$, is employed for non-linearity. $c_i$, $\alpha_{ij}$ and $\beta_{ij}$ are trainable parameters. More details about KAN are provided in Liu et al[2].

**Text S4: Model Training and Hyperparameters**
In δHBV1.0, LSTM generates time series of all parameters for HBV model, and the static parameters only used the prediction of the last step. In δHBV2.0, the LSTM and MLP models are trained jointly with the HBV modules, where the LSTM is used for dynamic parameterization and the MLP for static parameterization. In addition, a single LSTM model was also trained separately as a purely data-driven benchmark model for streamflow prediction. All models in this work used AdamDelta optimizer to automatically adjust the learning rate. LSTM has a hidden size of 256 with a dropout rate of 0.5. MLP has a hidden size of 4096 with a dropout rate of 0.5. The large hidden size for MLP is due to the extensive number of static and routing parameters, its simple structure, and the vast number of MERIT basins[1]. δHBVs were trained to 100 epochs with a batch size of 100. The purely data-driven LSTMs were trained to 300 with a same bath size of δHBVs. δMC used two KANs to estimate the hydraulic parameters and leakage parameters for MC, respectively. KANs used a hidden size of 8. δMC was trained for only 5 epochs with a mini-batch size of 64 basins as its input, as the runoff simulation from δHBV2.0 is already well-trained.